\documentclass[trackchanges,twocolumn]{aastex701}

\shorttitle{DESI PV Survey DR1: zero-point and $H_0$}
\shortauthors{Carr et al.}

\usepackage{comment}
\usepackage{booktabs}
\usepackage{amsmath}

\newcommand{\kmsMpc}{\ensuremath{\text{km}~\text{s}^{-1}~\text{Mpc}^{-1}}}
\newcommand{\hMpc}{$h^{-1}$Mpc}
\newcommand{\nSNlc}{137}
\newcommand{\nSN}{93}
\newcommand{\HSN}{73.7}
\newcommand{\HstatSN}{0.06}
\newcommand{\HsystSN}{0.29}
\newcommand{\HtotsystSN}{1.1}
\newcommand{\HtotSN}{1.1}
\newcommand{\nSBF}{27}
\newcommand{\HSBF}{74.1}
\newcommand{\HstatSBF}{0.06}
\newcommand{\nSNSBF}{109}
\newcommand{\HSNSBF}{73.9}
\newcommand{\HstatSNSBF}{0.06}

\begin{document}

\title{The DESI DR1 Peculiar Velocity Survey: global zero-point and $H_0$ constraints}
\correspondingauthor{A.~Carr}

\author[0000-0003-4074-5659, gname='Anthony', sname='Carr']{A.~Carr}
\affiliation{Korea Astronomy and Space Science Institute, 776 Daedeokdae-ro, Yuseong-gu, Daejeon 34055, Republic of Korea}
\email{anthonycarr@kasi.re.kr}

\author[0000-0002-1081-9410, gname='Cullan', sname='Howlett']{C.~Howlett}
\affiliation{School of Mathematics and Physics, University of Queensland, Brisbane, QLD 4072, Australia}
\email{c.howlett@uq.edu.au}

\author[0000-0003-3433-2698, gname='Ariel', sname='Amsellem']{A.~J.~Amsellem}
\affiliation{Department of Physics, Carnegie Mellon University, 5000 Forbes Avenue, Pittsburgh, PA 15213, USA}
\email{aamselle@andrew.cmu.edu}

\author[0000-0002-4213-8783, gname='Tamara', sname='Davis']{Tamara~M.~Davis}
\affiliation{School of Mathematics and Physics, University of Queensland, Brisbane, QLD 4072, Australia}
\email{tamarad@physics.uq.edu.au}

\author[0000-0002-1809-6325, gname='Khaled', sname='Said']{K.~Said}
\affiliation{School of Mathematics and Physics, University of Queensland, Brisbane, QLD 4072, Australia}
\email{k.saidahmedsoliman@uq.edu.au}

\author[0000-0002-7464-2351, gname='David', sname='Parkinson']{D.~Parkinson}
\affiliation{Korea Astronomy and Space Science Institute, 776 Daedeokdae-ro, Yuseong-gu, Daejeon 34055, Republic of Korea}
\email{davidparkinson@kasi.re.kr}

\author[0000-0002-6011-0530, gname='Antonella', sname='Palmese']{A.~Palmese}
\affiliation{Department of Physics, Carnegie Mellon University, 5000 Forbes Avenue, Pittsburgh, PA 15213, USA}
\email{apalmese@andrew.cmu.edu}

\author[gname='Jessica Nicole', sname='Aguilar']{J.~Aguilar}
\affiliation{Lawrence Berkeley National Laboratory, 1 Cyclotron Road, Berkeley, CA 94720, USA}
\email{jaguilar@lbl.gov}

\author[0000-0001-6098-7247, gname='Steven', sname='Ahlen']{S.~Ahlen}
\affiliation{Department of Physics, Boston University, 590 Commonwealth Avenue, Boston, MA 02215 USA}
\email{ahlen@bu.edu}

\author[gname='Julian', sname='Bautista']{J.~Bautista}
\affiliation{Aix Marseille Univ, CNRS/IN2P3, CPPM, Marseille, France}
\email{bautista@cppm.in2p3.fr}

\author[0000-0001-5537-4710, gname='Segev', sname='BenZvi']{S.~BenZvi}
\affiliation{Department of Physics \& Astronomy, University of Rochester, 206 Bausch and Lomb Hall, P.O. Box 270171, Rochester, NY 14627-0171, USA}
\email{sbenzvi@ur.rochester.edu}

\author[0000-0001-9712-0006, gname='Davide', sname='Bianchi']{D.~Bianchi}
\affiliation{Dipartimento di Fisica ``Aldo Pontremoli'', Universit\`a degli Studi di Milano, Via Celoria 16, I-20133 Milano, Italy}
\affiliation{INAF-Osservatorio Astronomico di Brera, Via Brera 28, 20122 Milano, Italy}
\email{davide.bianchi1@unimi.it}

\author[0000-0002-5423-5919, gname='Chris', sname='Blake']{C.~Blake}
\affiliation{Centre for Astrophysics \& Supercomputing, Swinburne University of Technology, P.O. Box 218, Hawthorn, VIC 3122, Australia}
\email{cblake@astro.swin.edu.au}

\author[gname='David', sname='Brooks']{D.~Brooks}
\affiliation{Department of Physics \& Astronomy, University College London, Gower Street, London, WC1E 6BT, UK}
\email{david.brooks@ucl.ac.uk}

\author[gname='Todd', sname='Claybaugh']{T.~Claybaugh}
\affiliation{Lawrence Berkeley National Laboratory, 1 Cyclotron Road, Berkeley, CA 94720, USA}
\email{tmclaybaugh@lbl.gov}

\author[0000-0002-2169-0595, gname='Andrei', sname='Cuceu']{A.~Cuceu}
\affiliation{Lawrence Berkeley National Laboratory, 1 Cyclotron Road, Berkeley, CA 94720, USA}
\email{acuceu@lbl.gov}

\author[0000-0002-1769-1640, gname='Axel ', sname='de la Macorra']{A.~de la Macorra}
\affiliation{Instituto de F\'{\i}sica, Universidad Nacional Aut\'{o}noma de M\'{e}xico,  Circuito de la Investigaci\'{o}n Cient\'{\i}fica, Ciudad Universitaria, Cd. de M\'{e}xico  C.~P.~04510,  M\'{e}xico}
\email{macorra@fisica.unam.mx}

\author[gname='Peter', sname='Doel']{P.~Doel}
\affiliation{Department of Physics \& Astronomy, University College London, Gower Street, London, WC1E 6BT, UK}
\email{apd@star.ucl.ac.uk}

\author[0000-0002-9540-546X, gname='Kelly', sname='Douglass']{K.~Douglass}
\affiliation{Department of Physics \& Astronomy, University of Rochester, 206 Bausch and Lomb Hall, P.O. Box 270171, Rochester, NY 14627-0171, USA}
\email{kellyadouglass@rochester.edu}

\author[0000-0003-4992-7854, gname='Simone', sname='Ferraro']{S.~Ferraro}
\affiliation{Lawrence Berkeley National Laboratory, 1 Cyclotron Road, Berkeley, CA 94720, USA}
\affiliation{University of California, Berkeley, 110 Sproul Hall \#5800 Berkeley, CA 94720, USA}
\email{sferraro@lbl.gov}

\author[0000-0002-2890-3725, gname='Jaime E.', sname='Forero-Romero']{J.~E.~Forero-Romero}
\affiliation{Departamento de F\'isica, Universidad de los Andes, Cra. 1 No. 18A-10, Edificio Ip, CP 111711, Bogot\'a, Colombia}
\affiliation{Observatorio Astron\'omico, Universidad de los Andes, Cra. 1 No. 18A-10, Edificio H, CP 111711 Bogot\'a, Colombia}
\email{je.forero@uniandes.edu.co}

\author[0000-0001-9632-0815, gname='Enrique', sname='Gaztañaga']{E.~Gaztañaga}
\affiliation{Institut d'Estudis Espacials de Catalunya (IEEC), c/ Esteve Terradas 1, Edifici RDIT, Campus PMT-UPC, 08860 Castelldefels, Spain}
\affiliation{Institute of Cosmology and Gravitation, University of Portsmouth, Dennis Sciama Building, Portsmouth, PO1 3FX, UK}
\affiliation{Institute of Space Sciences, ICE-CSIC, Campus UAB, Carrer de Can Magrans s/n, 08913 Bellaterra, Barcelona, Spain}
\email{gaztanaga@gmail.com}

\author[0000-0003-3142-233X, gname='Satya ', sname='Gontcho A Gontcho']{S.~Gontcho A Gontcho}
\affiliation{Lawrence Berkeley National Laboratory, 1 Cyclotron Road, Berkeley, CA 94720, USA}
\affiliation{University of Virginia, Department of Astronomy, Charlottesville, VA 22904, USA}
\email{satya@virginia.edu}

\author[gname='Gaston', sname='Gutierrez']{G.~Gutierrez}
\affiliation{Fermi National Accelerator Laboratory, PO Box 500, Batavia, IL 60510, USA}
\email{gaston@fnal.gov}

\author[0000-0002-9136-9609, gname='Hiram K.', sname='Herrera-Alcantar']{H.~K.~Herrera-Alcantar}
\affiliation{Institut d'Astrophysique de Paris. 98 bis boulevard Arago. 75014 Paris, France}
\affiliation{IRFU, CEA, Universit\'{e} Paris-Saclay, F-91191 Gif-sur-Yvette, France}
\email{herreraa@iap.fr}

\author[0000-0002-6550-2023, gname='Klaus', sname='Honscheid']{K.~Honscheid}
\affiliation{Center for Cosmology and AstroParticle Physics, The Ohio State University, 191 West Woodruff Avenue, Columbus, OH 43210, USA}
\affiliation{Department of Physics, The Ohio State University, 191 West Woodruff Avenue, Columbus, OH 43210, USA}
\affiliation{The Ohio State University, Columbus, 43210 OH, USA}
\email{kh@physics.osu.edu}

\author[0000-0001-6558-0112, gname='Dragan', sname='Huterer']{D.~Huterer}
\affiliation{Department of Physics, University of Michigan, 450 Church Street, Ann Arbor, MI 48109, USA}
\affiliation{University of Michigan, 500 S. State Street, Ann Arbor, MI 48109, USA}
\email{huterer@umich.edu}

\author[0000-0002-6024-466X, gname='Mustapha', sname='Ishak']{M.~Ishak}
\affiliation{Department of Physics, The University of Texas at Dallas, 800 W. Campbell Rd., Richardson, TX 75080, USA}
\email{mishak@utdallas.edu}

\author[0000-0003-0201-5241, gname='Dick', sname='Joyce']{R.~Joyce}
\affiliation{NSF NOIRLab, 950 N. Cherry Ave., Tucson, AZ 85719, USA}
\email{richard.joyce@noirlab.edu}

\author[0000-0001-6315-8743, gname='Alex G.', sname='Kim']{A.~G.~Kim}
\affiliation{Lawrence Berkeley National Laboratory, 1 Cyclotron Road, Berkeley, CA 94720, USA}
\email{agkim@lbl.gov}

\author[0000-0002-8828-5463, gname='David', sname='Kirkby']{D.~Kirkby}
\affiliation{Department of Physics and Astronomy, University of California, Irvine, 92697, USA}
\email{dkirkby@uci.edu}

\author[0000-0001-6356-7424, gname='Anthony', sname='Kremin']{A.~Kremin}
\affiliation{Lawrence Berkeley National Laboratory, 1 Cyclotron Road, Berkeley, CA 94720, USA}
\email{akremin@lbl.gov}

\author[gname='Ofer', sname='Lahav']{O.~Lahav}
\affiliation{Department of Physics \& Astronomy, University College London, Gower Street, London, WC1E 6BT, UK}
\email{o.lahav@ucl.ac.uk}

\author[0000-0002-6731-9329, gname='Claire', sname='Lamman']{C.~Lamman}
\affiliation{The Ohio State University, Columbus, 43210 OH, USA}
\email{lamman.1@osu.edu}

\author[0000-0003-1838-8528, gname='Martin', sname='Landriau']{M.~Landriau}
\affiliation{Lawrence Berkeley National Laboratory, 1 Cyclotron Road, Berkeley, CA 94720, USA}
\email{mlandriau@lbl.gov}

\author[0000-0001-7178-8868, gname='Laurent', sname='Le Guillou']{L.~Le~Guillou}
\affiliation{Sorbonne Universit\'{e}, CNRS/IN2P3, Laboratoire de Physique Nucl\'{e}aire et de Hautes Energies (LPNHE), FR-75005 Paris, France}
\email{llg@lpnhe.in2p3.fr}

\author[0000-0003-1887-1018, gname='Michael', sname='Levi']{M.~E.~Levi}
\affiliation{Lawrence Berkeley National Laboratory, 1 Cyclotron Road, Berkeley, CA 94720, USA}
\email{melevi@lbl.gov}

\author[0000-0003-4962-8934, gname='Marc', sname='Manera']{M.~Manera}
\affiliation{Departament de F\'{i}sica, Serra H\'{u}nter, Universitat Aut\`{o}noma de Barcelona, 08193 Bellaterra (Barcelona), Spain}
\affiliation{Institut de F\'{i}sica d’Altes Energies (IFAE), The Barcelona Institute of Science and Technology, Edifici Cn, Campus UAB, 08193, Bellaterra (Barcelona), Spain}
\email{mmanera@ifae.es}

\author[0000-0002-1125-7384, gname='Aaron', sname='Meisner']{A.~Meisner}
\affiliation{NSF NOIRLab, 950 N. Cherry Ave., Tucson, AZ 85719, USA}
\email{aaron.meisner@noirlab.edu}

\author[gname='Ramon', sname='Miquel']{R.~Miquel}
\affiliation{Instituci\'{o} Catalana de Recerca i Estudis Avan\c{c}ats, Passeig de Llu\'{\i}s Companys, 23, 08010 Barcelona, Spain}
\affiliation{Institut de F\'{i}sica d’Altes Energies (IFAE), The Barcelona Institute of Science and Technology, Edifici Cn, Campus UAB, 08193, Bellaterra (Barcelona), Spain}
\email{rmiquel@ifae.es}

\author[0000-0002-2733-4559, gname='John', sname='Moustakas']{J.~Moustakas}
\affiliation{Department of Physics and Astronomy, Siena University, 515 Loudon Road, Loudonville, NY 12211, USA}
\email{jmoustakas@siena.edu}

\author[0000-0001-9070-3102, gname='Seshadri', sname='Nadathur']{S.~Nadathur}
\affiliation{Institute of Cosmology and Gravitation, University of Portsmouth, Dennis Sciama Building, Portsmouth, PO1 3FX, UK}
\email{seshadri.nadathur@port.ac.uk}

\author[0000-0002-0644-5727, gname='Will', sname='Percival']{W.~J.~Percival}
\affiliation{Department of Physics and Astronomy, University of Waterloo, 200 University Ave W, Waterloo, ON N2L 3G1, Canada}
\affiliation{Perimeter Institute for Theoretical Physics, 31 Caroline St. North, Waterloo, ON N2L 2Y5, Canada}
\affiliation{Waterloo Centre for Astrophysics, University of Waterloo, 200 University Ave W, Waterloo, ON N2L 3G1, Canada}
\email{will.percival@uwaterloo.ca}

\author[0000-0001-7145-8674, gname='Francisco', sname='Prada']{F.~Prada}
\affiliation{Instituto de Astrof\'{i}sica de Andaluc\'{i}a (CSIC), Glorieta de la Astronom\'{i}a, s/n, E-18008 Granada, Spain}
\email{fprada@iaa.es}

\author[0000-0001-6979-0125, gname='Ignasi', sname='Pérez-Ràfols']{I.~P\'erez-R\`afols}
\affiliation{Departament de F\'isica, EEBE, Universitat Polit\`ecnica de Catalunya, c/Eduard Maristany 10, 08930 Barcelona, Spain}
\email{ignasi.perez.rafols@upc.edu}

\author[0000-0001-7950-7864, gname='Fei', sname='Qin']{F.~Qin}
\affiliation{Aix Marseille Univ, CNRS/IN2P3, CPPM, Marseille, France}
\email{qin@cppm.in2p3.fr}

\author[0009-0003-4767-9794, gname='Caitlin', sname='Ross']{C.~Ross}
\affiliation{School of Mathematics and Physics, University of Queensland, Brisbane, QLD 4072, Australia}
\email{c.ross1@uq.net.au}

\author[gname='Graziano', sname='Rossi']{G.~Rossi}
\affiliation{Department of Physics and Astronomy, Sejong University, 209 Neungdong-ro, Gwangjin-gu, Seoul 05006, Republic of Korea}
\email{graziano@sejong.ac.kr}

\author[0000-0002-9646-8198, gname='Eusebio', sname='Sanchez']{E.~Sanchez}
\affiliation{CIEMAT, Avenida Complutense 40, E-28040 Madrid, Spain}
\email{eusebio.sanchez@ciemat.es}

\author[gname='David', sname='Schlegel']{D.~Schlegel}
\affiliation{Lawrence Berkeley National Laboratory, 1 Cyclotron Road, Berkeley, CA 94720, USA}
\email{djschlegel@lbl.gov}

\author[0000-0002-6588-3508, gname='Hee-Jong', sname='Seo']{H.~Seo}
\affiliation{Department of Physics \& Astronomy, Ohio University, 139 University Terrace, Athens, OH 45701, USA}
\email{seoh@ohio.edu}

\author[gname='David', sname='Sprayberry']{D.~Sprayberry}
\affiliation{NSF NOIRLab, 950 N. Cherry Ave., Tucson, AZ 85719, USA}
\email{david.sprayberry@noirlab.edu}

\author[0000-0003-1704-0781, gname='Gregory', sname='Tarlé']{G.~Tarl\'{e}}
\affiliation{University of Michigan, 500 S. State Street, Ann Arbor, MI 48109, USA}
\email{gtarle@umich.edu}

\author[0000-0002-7638-2880, gname='Ryan', sname='Turner']{R. J.~Turner}
\affiliation{Centre for Astrophysics \& Supercomputing, Swinburne University of Technology, P.O. Box 218, Hawthorn, VIC 3122, Australia}
\email{rjturner@swin.edu.au}

\author[gname='Benjamin Alan', sname='Weaver']{B.~A.~Weaver}
\affiliation{NSF NOIRLab, 950 N. Cherry Ave., Tucson, AZ 85719, USA}
\email{benjamin.weaver@noirlab.edu}

\author[0000-0002-7305-9578, gname='Pauline', sname='Zarrouk']{P.~Zarrouk}
\affiliation{Sorbonne Universit\'{e}, CNRS/IN2P3, Laboratoire de Physique Nucl\'{e}aire et de Hautes Energies (LPNHE), FR-75005 Paris, France}
\email{pauline.zarrouk@lpnhe.in2p3.fr}

\author[0000-0001-5381-4372, gname='Rongpu', sname='Zhou']{R.~Zhou}
\affiliation{Lawrence Berkeley National Laboratory, 1 Cyclotron Road, Berkeley, CA 94720, USA}
\email{rongpuzhou@lbl.gov}

\author[0000-0002-6684-3997, gname='Hu', sname='Zou']{H.~Zou}
\affiliation{National Astronomical Observatories, Chinese Academy of Sciences, A20 Datun Road, Chaoyang District, Beijing, 100101, P.~R.~China}
\email{zouhu@nao.cas.cn}

\begin{abstract}
The Dark Energy Spectroscopic Instrument (DESI) in its first Data Release (DR1) already provides more than 100,000 galaxies with relative distance measurements. The primary purpose of this paper is to perform the calibration of the zero-point for the DESI Fundamental Plane and Tully-Fisher relations, which allows us to measure the Hubble constant, $H_0$. This sample has a lower statistical uncertainty than any previously used to measure $H_0$, and we investigate the systematic uncertainties in absolute calibration that could limit the accuracy of that measurement.  
We improve upon the DESI Early Data Release Fundamental Plane $H_0$ measurement by a) using a group catalog to increase the number of calibrator galaxies and b) investigating alternative calibrators in the nearby Universe.  
Our baseline measurement calibrates to the SH0ES/Pantheon+ type Ia supernovae, and finds $H_0=\HSN\pm\HstatSN\;(\text{stat.})\pm\HtotsystSN\;(\text{syst.})$ \kmsMpc{}. Calibrating to surface brightness fluctuation (SBF) distances yields a similar $H_0$.  We explore measurements using other calibrators, but these are currently less precise since the overlap with DESI peculiar velocity tracers is much smaller. 
In future data releases with an even larger peculiar velocity sample, we plan to calibrate directly to Cepheids and the tip of the red giant branch, which will enable the uncertainty to decrease towards a percent-level measurement of $H_0$. This will provide an alternative to supernovae as the Hubble flow sample for $H_0$ measurements. 
\end{abstract}


\section{Introduction}
Distances to extragalactic objects provide the independent information, beyond redshifts, needed to understand the expansion of the Universe. Redshifts are generated not only by the homogeneous expansion (recession velocities) but also by motions induced by local gravitational inhomogeneities (peculiar velocities).  These two velocities cannot be disentangled using redshifts alone. Accurate distance measurements break this degeneracy, enabling tests of the Hubble–Lemaître law and studies of the local galaxy distribution through peculiar velocities.

While galaxy redshift surveys have provided an ever-increasing number of precise spectroscopic redshifts, distances are more difficult to measure. Peculiar velocity (PV) samples have therefore always been much smaller in size than redshift samples, even over the same redshift range. However, as the survey technology has improved, the number of peculiar velocity measurements has also grown. In a similar fashion to other statistical methods in cosmology, current and future nearby galaxy surveys will enable PV science to become precision cosmology.

One such ongoing effort is the Dark Energy Spectroscopic Instrument survey, which is nearing 5 years of observation covering most of the northern hemisphere sky.  DESI is able to measure nearly 5,000 redshifts simultaneously and is performing a large redshift survey in order to measure the large-scale structure of the Universe and the properties of dark energy.  Simultaneously, it is running a dedicated program to measure peculiar velocities in the nearby ($z<0.1$) Universe, which will precisely measure the growth rate of structure and deliver the largest PV sample to date.

To achieve this, we use DESI spectra to measure the redshifts of and direct {\em distances} to galaxies, which can be converted into peculiar velocities. 
This will be achieved by targeting two well-established distance indicators, namely spiral and elliptical galaxies that follow the Tully-Fisher \citep[TF;][]{Tully1977} and Fundamental Plane \citep[FP;][]{Djorgovski1987, Dressler1987} relations, respectively; we introduce these techniques in Sec.~\ref{sec:DESI_PV_DR1}.

DESI was forecast to measure the distance/PV of around 133,000 elliptical and 53,000 spiral galaxies over its initial 14,000 square degree five-year observation program \citep{SurveyOps.Schlafly.2023,Saulder2023}, which is significantly larger than all other current PV catalogs to date.
However, these figures are set to increase beyond the forecast, given that DESI has been extended to an eight-year survey covering 17,000 square degrees.

This paper is one in a series of DESI papers that present DESI PV measurements and use them to constrain cosmology (see Sec~\ref{sec:DESI_PV_DR1}).   While one of the primary foci of the DESI PV survey is to measure the growth of structure through the correlations of PVs, in this paper, we focus on measuring the recent expansion history of the Universe, $H_0$.

The TF and FP measurements provide distances that, when combined with redshift information, result in an alternative Hubble diagram to the more common three-rung SN distance ladder. 
However, the TF and FP relations only provide relative distance indications, much the same as SNe Ia; they both require a zero-point that must be tied to external measurements to be used on an absolute scale.

While the correlations between velocities are insensitive to the distance zero-point, the expansion rate is fully degenerate with the zero-point.
Without an accurate global zero-point, not only would the PVs have the wrong magnitude, but the TF and FP Hubble diagrams may exhibit a relative offset, which would translate to unphysical redshift evolution, which could be interpreted as bulk flows or higher-order velocity moments.

Fundamental plane measurements from the DESI Early Data Release (EDR) have already been used to measure $H_0$ \citep{Said2025}, by using the distance to the Coma cluster measured using surface brightness fluctuations \citep[SBF;][]{Jensen2021} to calibrate the FP relation.  Unfortunately, that measurement suffers from a large statistical error due to using a single cluster as the calibration anchor.  Here we improve on that early work by calibrating with type Ia supernovae (SNe Ia) as the absolute distance measure.  This has the downside of creating a four-rung distance ladder, as the SNe are the third rung of the typical distance ladder, which means the calibration and any systematic errors depend on the previous rungs and are inherited by the DESI distances.  However, it enables us to reduce the calibration uncertainty significantly because we can use many anchors. 

Ideally, we would like to calibrate using fewer rungs of the distance ladder, such as directly to Cepheid variables, masers, or the tip of the red giant branch (TRBG).  However, even though the first data release of DESI (DR1) provides the largest peculiar velocity sample to date, the overlap between DESI PV galaxies and primary calibrators is very small. Therefore, currently, supernovae remain as our best anchor due to their superior statistics. 

To increase the number of SN Ia host galaxies that overlap with the DESI PV sample, we use the established technique of matching not just galaxies, but also galaxy groups.
Galaxy groups have been used to reduce dispersion on the SN Hubble diagram caused by the peculiar velocity dispersion of individual host galaxies within clusters (see, \textit{e.g.}, \citealt{Leget2018,Peterson2022,Peterson2025}), which is similar to our goal of accurate calibration. 
The EDR FP analysis used a single distance to the Coma cluster applied to all $\sim1600$ EDR FP galaxies in Coma \citep{Said2025}, and we simply expand that idea to using all available groups/clusters in the DR1 footprint.
Similar to their use in SN cosmology, this reduces the dispersion in calibration caused by too few calibrators.
By assuming that all group members are at the same distance (more valid as redshift increases), we approximately double the number of matches with the DESI PV sample, and thus double the number of calibrators (see Sec.~\ref{sec:galaxy_groups}).  

This paper is set out as follows: in Sec.~\ref{sec:DESI_PV_DR1} we discuss in more detail the first data release of the DESI Peculiar Velocity Survey, then in Sec.~\ref{sec:calibrators} we discuss the calibrators we use to measure $H_0$.
In Sec.~\ref{sec:galaxy_groups}, we explain the method we employ of using groups of galaxies to increase the overlap of the DR1 sample with our calibrators, then we detail the process for finding the global zero-point in Sec.~\ref{sec:zp_process}. We then discuss the constraints on $H_0$ we obtain from each calibrator in Sec.~\ref{sec:cosmology} and finally summarize our findings in Sec.~\ref{sec:summary}.

\section{DESI and the DESI Peculiar Velocity Survey}\label{sec:DESI_PV_DR1}
\subsection{The Dark Energy Spectroscopic Instrument}
DESI is a spectroscopic instrument mounted on the Mayall 4-meter telescope at Kitt Peak National Observatory that can observe almost 5,000 objects simultaneously over a ${\sim}3^{\circ}$-diameter field-of-view thanks to the multiplexed robotic fiber system \citep{DESI2016b.Instr,DESI2022.KP1.Instr,FocalPlane.Silber.2023,Corrector.Miller.2023,FiberSystem.Poppett.2024}.
Each of the fibers can move independently within their own `patrol radius' of 1.48 arcmin on the sky, and are 1.5 arcsec in diameter.
The fibers feed 10 spectrographs that cover the near-UV to near-infrared wavelength range (3,600 to 9,800 \AA) with a spectral resolution ranging from 2,000 in the blue camera to 5,000 in the red camera \citep{Spectro.Pipeline.Guy.2023}.
The full survey \citep{DESI2016a.Science} will observe over 60 million galaxies and quasars over eight years. 

The first data release of spectra, redshifts and value-added catalogs is already public \citep{DESI2024.I.DR1}.
This covers the first year of DESI observations over an area of almost 10,000 square degrees to varying depth.
In addition, the cosmological results from the first and second data release, such as redshift space distortions and baryon acoustic oscillation measurements \citep[see, for example][]{DESI2024.VII.KP7B,DESI.DR2.BAO.cosmo} are also public.
What we describe in the current series of papers is the first data release of the Peculiar Velocity Survey specifically.

\subsection{The DESI peculiar velocity program}

The DESI PV program uses two distance indicators to derive peculiar velocities: the Tully-Fisher and Fundamental Plane techniques.  The TF method uses the rotation velocity of spiral galaxies to estimate their true absolute magnitude, making them standard candles.  The FP method uses the central velocity dispersion and mean surface brightness of elliptical galaxies to estimate their true size, making them standard rulers. 

DESI uses a novel technique to measure TF distances, by placing at least two spectroscopic fibers along the axis of rotation of spiral galaxies.  DESI is thus able to measure their rotation velocities directly from the redshifts, which enables the measurement of the Tully-Fisher relation.  This way of measuring rotation velocity differs from the typical method using the width of H\textsc{i} 21 cm emission; see \citet{Saulder2023} and \citet{Douglass2025} for more detail.

To measure FP distances, DESI places a fiber at the core of elliptical galaxies to measure their central velocity dispersion.  Accompanied by photometric measurements of surface brightness, this enables the measurement of the Fundamental Plane relation; see \citet{Saulder2023} and \citet{Said2025} for more detail.

Neither relation is complete without accompanying photometry, which is supplied by the DESI Legacy Imaging Survey \citep{Dey2019}.  This is made up of three surveys: the Dark Energy Camera Legacy Survey (DECaLS), the Beijing-Arizona Sky Survey (BASS; $g$- and $r$-bands only) and the Mayall $z$-band Legacy Survey (MzLS).

The primary focus of DR1 is to constrain the growth rate of structure at low redshift by harnessing the power of PVs.
This goal is achieved in four ways: by correlating supernova Hubble residuals with large-scale structure along the lines of sight to probe the local growth rate of structure \citep{DESI_DR1_PV_SN_xcorr}; by combining DESI BGS density field information with the TF and FP peculiar velocities using the maximum likelihood fields method \citep{DESI_DR1_PV_max_like}; by using density and velocity auto- and cross-correlations \citep{DESI_DR1_PV_dens_vel_corr}; and finally, by using density and momentum power spectra \citep{DESI_DR1_PV_power_spec}.

The construction and analysis of the DR1 TF sample is described in \citet{DESI_DR1_TF} and the FP sample in \citet{DESI_DR1_FP}, along with the mock catalogs used for validation and estimating the covariance between each $f\sigma_8$ measurement in \citet{DESI_DR1_PV_mocks}.

\subsection{DESI PV targeting and selection cuts}\label{sec:cuts}
The Peculiar Velocity Survey is a subset of the Bright Galaxy Survey \citep[BGS;][]{BGS.TS.Hahn.2023}, which targets all galaxies with $r$-band magnitude $<19.5$, plus additional color-selected galaxies in the range $19.5<r<20.175$ mag. The color cuts applied to the fainter sample of galaxies ensure a high redshift success despite being faint, and are included to increase the density of BGS. The resulting BGS comprises 6 million galaxies at $z<0.6$ out of the 16 million galaxies in DR1 \citep{DESI2024.I.DR1}.

The targeting strategies for FP and TF are necessarily different.
For FP, which requires only a spectroscopic fiber placed at the center of an elliptical galaxy, the sample comprises all galaxies that satisfy a simple set of criteria applied to the DR1 data. The main requirements of the FP sample are a successful central velocity dispersion measurement and an elliptical morphology, leaving a sample of 108,810 galaxies in the range $0.0033<z_{\text{obs}}<0.1$.  This is then further reduced to 96,758 FP distances after removing duplicate observations and outliers from the FP relation \citep{DESI_DR1_FP}.

By contrast, the TF survey selects targets from the Siena Galaxy Atlas \citep[SGA;][]{Moustakas2023} and requires extra fibers placed at 40\% of the 26th isophote radius ($0.4R_{26}$).
The SGA is a size-limited galaxy sample constructed from the same imaging surveys as DESI targets.
The particular galactocentric radius was chosen because it can reliably result in a successful redshift despite the lower signal-to-noise ratio, and also recover the maximum rotation velocity.
However, while this limited radius underestimates the total rotation velocity more than it overestimates, it has been shown that a) DESI very well reproduces the rotation velocity calculated at $0.4R_{26}$ from integral field observations of the same targets \citep{Saulder2023,Douglass2025,DESI_DR1_TF}, and b) the TF relation can be self-consistently calibrated at a given galactocentric radius \citep{Yegorova2007}.

The TF sample is restricted in ways the FP sample is not; it is a targeted survey, galaxies must be large enough to place at least two spectroscopic fibers, and at least two successful redshift measurements must be made, with at least one being at much lower surface brightness.
Despite these restrictions, the DESI TF sample will still be the largest sample of TF galaxies to date after the completion of DESI.  In DR1, the rotational velocity of 10,262 spiral galaxies is measured, which drops to 8,082 galaxies at $z>0.0166$ after removing likely dwarf galaxies and TF outliers (see \citealt{DESI_DR1_TF}).  When combining TF and FP to measure a global zero-point, we apply the tighter redshift cut of $z_{\text{obs}}<0.1$ to TF as well, reducing the sample to 7,277. 

For the EDR, both TF and FP zero-point calibrations were restricted to the Coma cluster.
TF was calibrated on two supernovae \citep{Stahl2021}, and FP on the distance measured to Coma by SBF by \citet{Jensen2021}.
The FP analysis serves as a precursor to the current analysis, and we make several improvements to the methodology in addition to the large increase in sample size.

As we detail in the next sections, we make use of a range of calibrators, including the original SBF measurement used by EDR FP, as well as a new calibration on type Ia supernovae. We also extend the possible calibrators by making use of every galaxy in groups that contain calibrators.

\section{Calibrators}\label{sec:calibrators}
In order to attach the TF and FP distances to a distance ladder, we need to find galaxies within our DESI sample that have known distances.  We use several existing distance catalogs to calibrate our data and compare the results, specifically: 
\begin{enumerate}
    \item SNe Ia from SH0ES/Pantheon+ (fiducial)
    \item Distance to Coma from:
    \begin{itemize}
        \item SBF \citep{Jensen2021}
        \item SN Ia \citep{Scolnic2025}
    \end{itemize}
    \item Other individual distances from the Extragalactic Distance Database (EDD; \citealt{Tully2009}):
    \begin{itemize}
        \item Masers 
        \item Cepheids 
        \item SBF 
    \end{itemize}
\end{enumerate}
Finally, since the SBF distances we use are also calibrated using Cepheids \citep[by taking advantage of clusters also containing Cepheid galaxies, similar to the method we employ in this work; see][]{Jensen2021}, they can be combined with the SNe Ia for a consistent zero-point.

\subsection{Calibrating with SNe Ia}
The supernova distance ladder is constructed from three rungs: geometric distances (parallax, masers, eclipsing binaries) which link to galaxies that host either Cepheids (as with SH0ES; \citealt{SH0ES22}) or tip of the red giant branch stars \citep[TRGB, as with Carnegie-Chicago Hubble Project;][]{Freedman2019}, which then link to galaxies that host SNe.
We use type Ia supernovae from the SH0ES/Pantheon+ supernova sample \citep{SH0ES22, Pplus_DR}.\footnote{The data release of these SNe can be found at \url{https://github.com/PantheonPlusSH0ES/DataRelease}.}
Since this distance ladder contains systematic uncertainties from all three rungs, our analysis cannot in principle give a $H_0$ constraint tighter than the SH0ES measurement. 
Therefore, to reduce systematics, a shorter ladder (more direct calibration) is preferred, but the SH0ES/Pantheon+ SNe are still useful because only a few galaxies in our current sample host lower-rung calibrators.

We make use of the full covariance matrix wherever possible, which contains statistical and systematic uncertainties propagated from the lower rungs of the distance ladder along with those from the SNe themselves.
We carefully propagate these uncertainties through our analysis as an irreducible floor within our own covariance matrix. 
Since there are cases of multiple SNe or multiple observations of the same SN in some galaxies, the off-diagonal covariance of these SNe is enhanced compared to SNe in different galaxies.
There are also more SN light curves than there are ``independent SN distances'' required to calibrate distances to galaxies; in general, we quote the number of galaxies with a known distance unless otherwise stated.

This supernova sample is our fiducial choice for calibration because it provides almost 100 independent distances to DESI PV galaxies, and thus provides a large number of calibrators to reduce statistical uncertainty. 
The propagation of full covariance also enables the most accurate estimation of uncertainty that is not available with other calibrators.

\subsection{Calibrating with Coma}
For the FP EDR, \citet{Said2025} used a single measurement of the distance to the Coma cluster to zero-point the FP Hubble diagram. 
The distance to Coma incorporated all optical Cosmicflows-4 \citep{Tully2023} SBF distances (425 measurements) supplemented by HST infrared SBF measurements from \cite{Jensen2021}.
As a form of validation, and to show the improvements offered by DR1 over EDR, we perform the same calibration using the same SBF measurement to Coma.

To add to this, a recent publication measured the distance to Coma with the highest precision to date, using Pantheon+, along with new SNe Ia \citep{Scolnic2025}.
We also adopt this measurement as a calibrator for comparison to our own SN Ia calibration.

\subsection{Calibrating with masers, Cepheids, and SBF}
We searched for other calibrators in The Extragalactic Distance Database (EDD; \citealt{Tully2009}), which is a comprehensive compilation of distances from calibrators such as masers, Cepheids, and TRGB, as well as distance tracers including SNe Ia, SNe II, SBF, TF, and FP measurements. 

The DR1 PV catalog currently has no overlap with TRGB galaxies, but we expect this to change with future data releases.
As we explain in Sec.~\ref{sec:zp_process}, we apply a redshift cut of $z>0.01$ when calculating the global zero-point, which removes the Cepheids from the sample of calibrators.  
This may also improve in future data releases if we can make use of the data below $z=0.01$ or if an increased sample size overlaps with higher redshift Cepheids.
To summarize, we show in Table~\ref{tab:cal_cuts} the calibrators (including SNe Ia) and how many remain a) in DR1 and b) after applying the $z>0.01$ cut.

\begin{deluxetable*}{cccccc}
\tablewidth{0pt}
\tablehead{ & Dist.~ladder & & & in DR1 & \vspace{-3pt}\\
\colhead{Calibrator} & \colhead{rung\tablenotemark{a}} & \colhead{Total} & \colhead{in DR1} & \&\;$z>0.01$
  & \colhead{Reference}}
\tablecaption{Number of calibrators remaining after DR1 cross-match and applying the redshift cut. \label{tab:cal_cuts}}
\startdata
SNe Ia\tablenotemark{b} & 3rd & 618 & 100 & \nSN{} & 1 \\
SBF & 3rd & 469 & 47 & \nSBF{} & 2--5 \\
Masers & 1st & 6 & 2 & 2 & 6, 7 \\
Cepheids & 2nd & 76 & 4 & 0 & 8--11 \\
TRGB & 2nd & 489 & 0 & 0 & 12 \\
\enddata
\tablenotetext{a}{The typical distance ladder rung the calibrator inhabits. Using the calibrator results in a distance ladder of this length plus one.}
\tablenotetext{b}{For SNe we consider only $z<0.1$.}
\tablereferences{(1) \citealt{Pplus_DR}; (2) \citealt{SBFTonry}; (3) \citealt{SBFBlakeslee}; (4) \citealt{SBFCantiello}; (5) \citealt{Jensen2021}; (6) \citealt{MASReid}; (7) \citealt{Pesce2020}; (8) \citealt{CEPHFreedman}; (9) \citealt{CEPHBhardwaj}; (10) \citealt{CEPHBentz}; (11) \citealt{SH0ES22}; (12) \citealt{TRGBAnand}.}
\end{deluxetable*}

\section{Distances to Galaxy groups}\label{sec:galaxy_groups}
DESI DR1 is relatively incomplete, so the overlap of the PV sample with the set of calibrator galaxies (galaxies with known distances) is currently small.
This will improve as DESI continues to observe; however, for now, we turn to galaxy groups to increase the number of matches to calibrator galaxies.
Taking advantage of galaxy groups allows for a more precise and accurate zero-point since we have access to a larger sample of calibrators, and we are able to average the distances of multiple calibrators within the same group.
The only requirement for calibrating a zero-point is to know the distance to a galaxy, so we assume all galaxies within a given group are at the same distance.
At very low redshift, this approximation begins to break down for large clusters, but the effects are negligible for calibrating, since most constraining power comes from larger redshifts. 

\begin{figure*}
    \centering
    \includegraphics[width=0.495\linewidth]{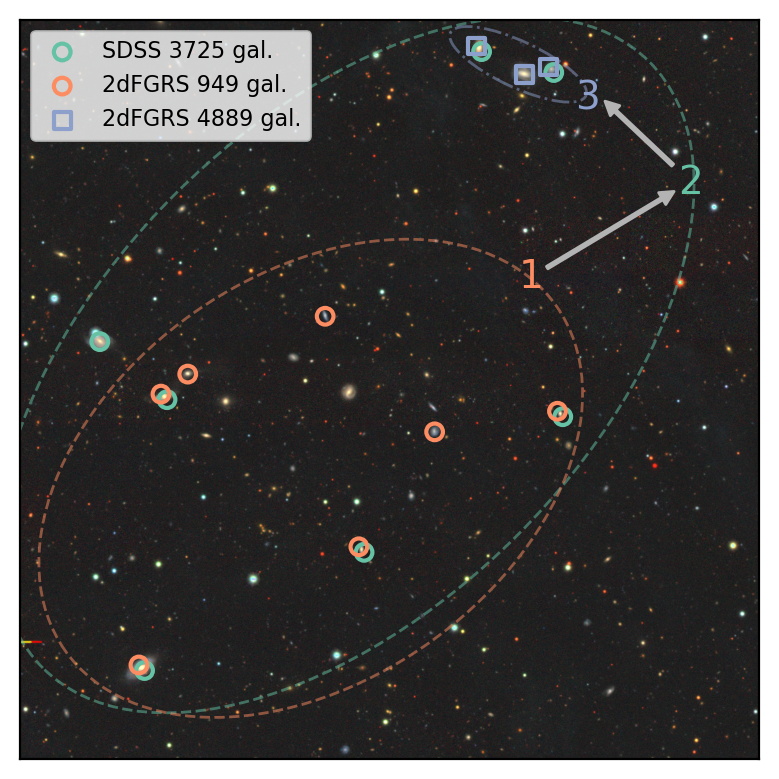}
    \includegraphics[width=0.495\linewidth]{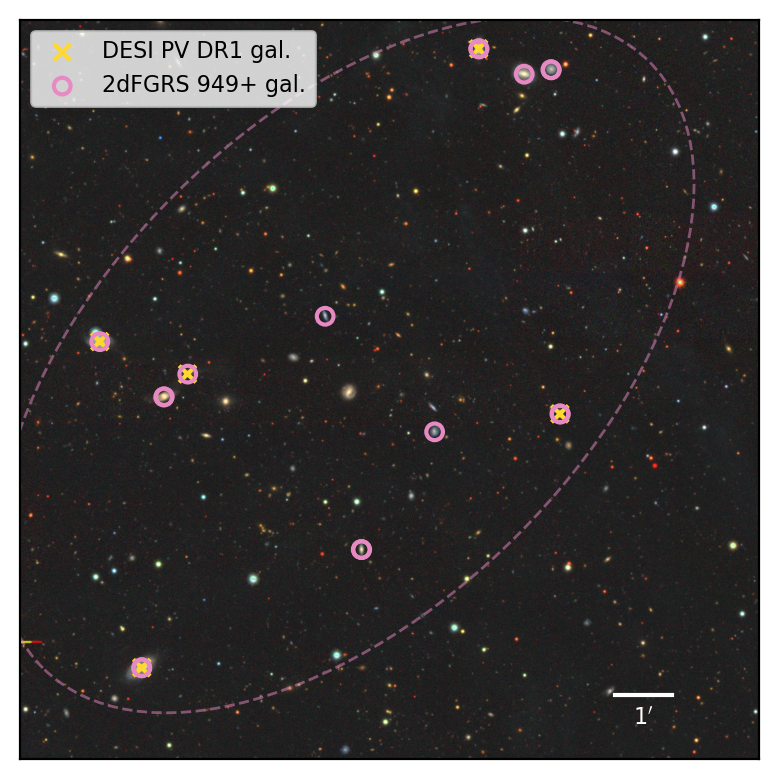}
    \caption{Example of how we combine group catalogs to discover calibrators that are in the same group as DESI PV galaxies. Taking Lim group 2dFGRS 949 (orange circles in the left panel) as the initial group that contains one of our DESI PV galaxies (step 3), the SDSS group (green circles) was found to overlap (step 4), and then 2dFGRS 4889 was linked (step 6). The final group is defined as all galaxies discovered in the process (pink circles in the right panel).  If any galaxy in the final group has a known distance, we can use it to calibrate the entire group and any PV galaxies in it (yellow crosses in the right panel). The background image is from the DESI Legacy Survey viewer, and the ellipses are for visualization only. To distinguish the galaxies that appear in multiple groups in the left panel, the individual symbols are offset from the center of the galaxy. The group identifiers are internal to the Lim catalog (a `+' denotes our expanded definition).}
    \label{fig:group_method}
\end{figure*}

At the time of writing, there exists no DESI-based group catalog, mainly due to the redshift incompleteness, so instead we use established, external galaxy catalogs.
We use the group catalogs of \citet{Lim2017}, wherein the authors apply their group-finding algorithms\footnote{We specifically make use of the group catalog based on luminosities and spectroscopic redshifts, rather than relying on photometric redshifts or galaxy mass estimates.} to four galaxy surveys: the Two Degree Field Galaxy Redshift Survey (2dFGRS), the Six-Degree Field Galaxy Survey (6dFGS), the Two Micron All-Sky Redshift Survey (2MRS) and the Sloan Digital Sky Survey (SDSS).
Since there is no combined analysis of the four surveys, and each has a different depth and sky area, we search for the same groups as defined in each survey and combine them as a way of further increasing the sample size.
As this is not guaranteed to converge for all groups (although it is very likely), we only perform one further iteration of this member-gathering process.

To clarify the process:
\begin{enumerate}
    \item We created a superset of the four group subcatalogs by \citet{Lim2017}. Each galaxy may thus have multiple identifiers. 
    \item Any galaxy belonging to the Coma cluster was removed; we replaced this definition of Coma directly with the DESI galaxies belonging to the more complete definition from \citet{Saulder2023}.
    \item We then performed a 5 arcsec coordinate cross-match with the DESI PV galaxies to identify DR1 groups.
    \item A second 5 arcsec cross-match was performed around each matched galaxy to find all associated group IDs from the subcatalogs, \textit{i.e.}\ every definition that contained a DR1 galaxy.
    \item Groups were then redefined as the union of every subcatalog definition from the previous step.
    \item We then performed a second iteration of steps 4 and 5, linking any more group IDs and galaxies associated with the original group. This step consolidates groups that share members in any of the subcatalogs.
    \item Finally, we generated a catalog of our expanded groups, assigning each modified group a new, unique ID.
\end{enumerate}

The grouping process is sketched out in Figure~\ref{fig:group_method}.
In the particular example shown, one can start with any of the three smaller groups and end up with the same final result, but it is most helpful to start with one of the 2dFGRS groups.
All three groups are linked by the end of step 6, resulting in the panel on the right, which contains all identified galaxies, and fills in galaxies that were missed in the SDSS grouping as preferred by 2dFGRS and vice-versa.

This method results in over 13,500 groups or pairs, and 784 of these groups contain at least one FP \textit{and} TF galaxy. This corresponds to 25\% of all DESI PV galaxies being associated with at least one other galaxy (around 25,500 of the 104,000 combined TF+FP sample) in the 78,000 galaxy combined \citealt{Lim2017} group catalog, implying an average group size of three members.
In the case of groups containing FP galaxies, 64.4\% of groups contain a single FP galaxy, 
21.2\% contain two FP galaxies, 6.6\% contain three, 3.0\% contain four, and the remaining 5\% contain five or more FP galaxies (with Coma containing 209).
From the full catalog, we calculated the redshift of each group as the simple average of all group members' redshifts.

As a simple example of how effective the group catalog is, the direct overlap of SNe~Ia with DR1 galaxies is four for the TF sample and 41 for the FP sample.  This increases to a total overlap of \nSN\ SN~Ia distances (from \nSNlc\ light-curves) to DESI PV targets once we also include groups. 
This represents a more than two-fold increase in the number of SN Ia distances, but a nearly 10-fold increase in the number of galaxies that have SN distances.
From 45 total TF and FP galaxies that host SNe Ia, we now have 571 galaxies that either host SNe Ia themselves or within their group.

The magnitude completeness of each of the four catalogs is different, and each is much shallower than the DESI BGS survey.
This means that many of the fainter DESI galaxies do not appear in the group catalogs. 
The current method is adequate for this DR1 analysis, but in future, a native DESI group catalog is required for coverage to the magnitude depth of the PV catalog for the most accurate zero-point.

\subsection{Comparison of galaxies in groups vs all galaxies}
\begin{figure}
    \centering
    \includegraphics[width=\linewidth]{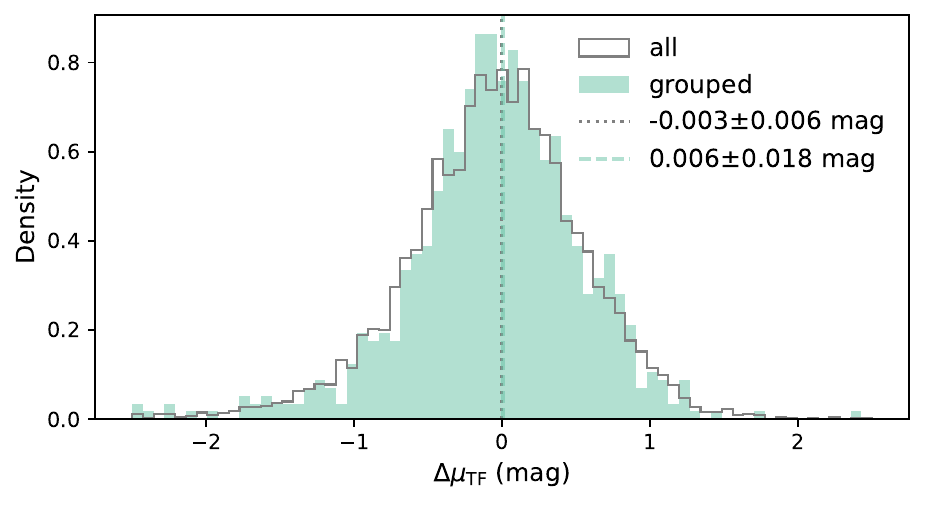}
    \caption{Distributions of TF distance indicators in groups (green) and in the full sample (gray). There is no evidence of a difference between the populations.}
    \label{fig:TF_group_corr}
\end{figure}

\begin{figure}
    \centering
    \includegraphics[width=\linewidth]{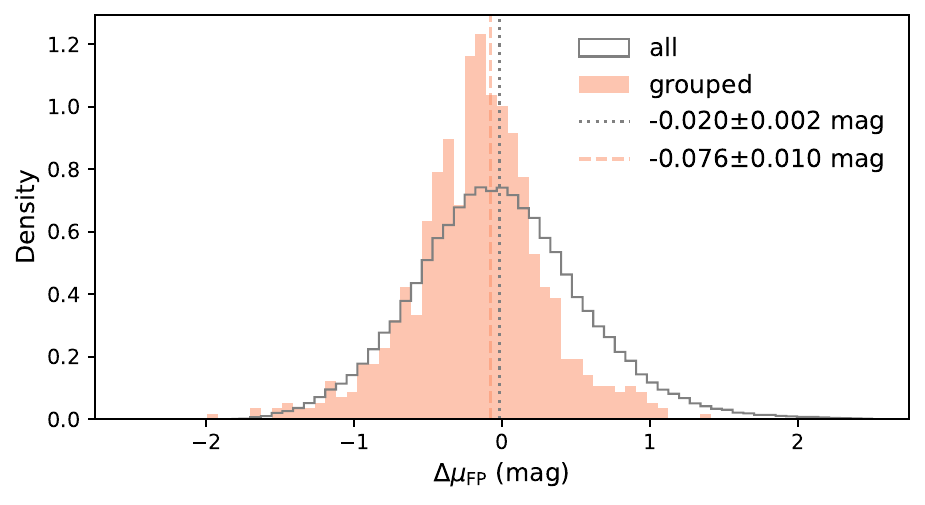}
    \caption{Distributions of FP distance indicators in groups (orange) and in the full sample (gray). The mean difference is significant enough to warrant correction back to the full sample mean.}
    \label{fig:FP_group_corr}
\end{figure}

To show how similar or different the DESI PV galaxies in groups are compared to the whole sample, we performed an inverse-variance-weighted two-sample Kolmogorov--Smirnov (K--S) test on the distances of the full and grouped populations for the FP and TF samples.
In all cases, weights are the inverse variance of the quantity being averaged, \textit{i.e.}\ $\langle x\rangle_Y=\sum_{i\in Y} w_ix_i$, where $x$ is usually a TF/FP distance, $w_i=1/\sigma^2_{x_i}$ and $Y$ is the subset being considered.
In addition to measurement uncertainties in each individual galaxy distance, the TF and FP relations both have an intrinsic dispersion which is also included in (and often dominates) the total uncertainty $\sigma_x$.
We expect that the population in groups (subset $G$) may be biased due to the group catalog having a shallower magnitude completeness.
To confirm or deny this, we compare the distribution of TF and FP distances to galaxies in groups and to all galaxies, in Figures~\ref{fig:TF_group_corr} and \ref{fig:FP_group_corr}.
The use of $\Delta\mu$ is expanded upon in Sec.~\ref{sec:zp_process}, but it can be viewed simply as distance for this test.
Basically, $\Delta\mu$ represents the apparent brightness (in magnitudes) compared to the prediction from a cosmological model, so a positive value means the galaxy is fainter, or further away than we expect.

We determined that the full TF and grouped TF distributions are drawn from the same distribution since the offset is consistent with 0, and K--S test resulted in a $p$-value of 0.16.
For FP, however, we find a statistically significant offset of 
\begin{equation}\label{eq:FP_mean_corr}
    \delta_{\text{FP}} = \langle\Delta\mu_{\text{FP}}\rangle_G-\langle\Delta\mu_{\text{FP}}\rangle = -0.057\pm 0.010 \text{ mag}, 
\end{equation}
with a $p$-value of 0.001.
The difference implies that the FP galaxies that were grouped in the \citet{Lim2017} catalogs are not drawn from the same distribution as the full population. 

The log-distance ratios of FP galaxies in larger groups appear to be systematically smaller than those in smaller groups \citep{Howlett2022}, so the DR1 FP analysis applies a richness correction \citep{DESI_DR1_FP}.
The richness correction is an empirical relation of possible physical origin found from comparing the mean log-distance ratio of FP galaxies in groups and field galaxies (as prescribed using a crude grouping algorithm, separate from the group catalog we use in this work); essentially, individual log-distance ratios are corrected to match the expected field galaxy value, based on the richness of its group.
Yet, even after this correction, we still see a statistically significant offset between the FP galaxies from groups and the full population.
The galaxies in groups are on average brighter, which is consistent with the theory that many fainter FP galaxies were missed in the group catalogs.
We thus corrected this offset for the grouped FP galaxies when calculating the weighted average of the FP sample, as not doing so would result in a biased full-sample zero-point.
We do \textit{not} apply a correction to the distances themselves; we only ensure the weighted average is representative of the full population.

\section{Zero-pointing process}\label{sec:zp_process}
The goal of the zero-pointing process is to anchor the FP and TF distance indicators to external distance data in a similar fashion to the SN distance ladder.
TF and FP fundamentally differ in that the TF relation provides an estimate of the galaxy's absolute \textit{magnitude} while the FP relation provides an estimate of the galaxy's absolute size. 

\subsection{The Tully-Fisher Relation}
The TF relation estimates a galaxy's absolute magnitude via 
\begin{equation}\label{eq:TF}
    M = a\log_{10}(V/V_0)+b,
\end{equation}
where $a$ and $b$ are the slope and intercept of the TF relation, $V$ is the galaxy's rotational velocity, and $V_0$ is the median of all rotational velocities in the sample. 

Thus, by comparing the absolute magnitude to apparent magnitude, the TF relation can be used to measure the luminosity distance $\bar{D}_L=(1+z_{\text{obs}})\bar{D}$. Here $\bar{D}=D(\bar{z})$ is the true comoving distance, defined at the cosmological redshift $\bar{z}$ (\textit{i.e.}\ the redshift with no PV contamination).\footnote{The use of the observed redshift in the factor of $(1+z_{\rm obs})$ arises because the reduction in photon energy and time dilation that give rise to this factor both depend on the observed redshift, not just the cosmological component \citep{Davis2019}.} 

Correspondingly, the natural distance measure for TF galaxies is the distance modulus, 
\begin{equation}\label{eq:mu_obs}
    \mu_{\text{obs}} = m - M = 5\log_{10}(\bar{D}_L/\text{Mpc}) + 25,
\end{equation}
which is a measure of the true luminosity distance, $\bar{D}_L$.
Note that both $m$ and $M$ (through rotation velocity $V$) already contain dust extinction and inclination corrections; for example, the observed magnitude of a spiral galaxy is more extincted at higher inclination due to more host-galaxy dust along the line-of-sight \citep{DESI_DR1_TF}.
The PV information is contained in the difference between $\mu_{\rm obs}$ and our model prediction for a distance modulus at the observed redshift,
\begin{equation}\label{eq:mu_mod}
    \mu_{\text{model}} = 5\log_{10}(D_L(z_{\text{CMB}},z_{\text{obs}})/\text{Mpc}) + 25.
\end{equation}
The CMB-frame redshift is as close as we can easily get to the true cosmological redshift --- it is obtained by correcting the observed redshift for the motion of our solar system. However, it still contains the source's PV contamination. By using $z_{\rm CMB}$ in the calculation for $D_L$, we will find a distance that is offset from the true $\bar{D}_L$:
\begin{equation}\label{eq:dmu}
    \Delta\mu =\mu_{\text{obs}}-\mu_{\text{model}}= 5\log_{10}(\bar{D}_L/D_L(z_{\text{CMB}},z_{\text{obs}})).
\end{equation}
That offset reveals how far $z_{\rm CMB}$ deviates from $\bar{z}$, which we attribute to the extra redshift given by a peculiar velocity through
\begin{equation}    \label{eq:zpec}
z_{\rm p}=(1+z_{\rm CMB})/(1+\bar{z})-1.
\end{equation}

\subsection{The Fundamental Plane Relation}
On the other hand, the FP measures the effective \textit{radius} of a galaxy via
\begin{equation}\label{eq:FP}
    \log_{10}(R_e) = a\log_{10}(\sigma_0) + b\log_{10}(I_e) + c,
\end{equation}
where $a$, $b$ and $c$ are the slopes and intercept of the FP relation, $\sigma_0$ is the central velocity dispersion and $I_e$ is the effective surface brightness.
As with the TF sample, the parameters in the FP relation have already had magnitude and velocity dispersion corrections applied; for example, through the same 1.5 arcsec diameter fiber, different physical proportions of a galaxy will be observed depending on its size and distance, so the observed dispersion $\sigma$ is corrected to a central dispersion $\sigma_0$ \citep{DESI_DR1_FP}.
Thus, by comparing the absolute size to apparent angular size $\theta$, the FP relation can be used to measure the angular diameter distance $D_A=R_e/\theta=\bar{D}/(1+z_{\text{obs}})$.   

The natural measurement for the FP is the log-distance ratio,
\begin{equation}\label{eq:eta}
    \eta = \log_{10}(D(z_{\text{CMB}})/\bar{D}),
\end{equation}
where the distances are comoving, as measured from physical galaxy sizes through Eq.~(\ref{eq:FP}).
Eq.~(\ref{eq:eta}) encapsulates the effects of peculiar velocities through the comparison of distances inferred from the FP relation (the ``true'' distance) and the model distance calculated at $z_{\text{CMB}}$.
Log-distance ratios are therefore a measure of the departure from the true distance arising because $z_{\rm CMB}\ne \bar{z}$, where the difference is due to the extra redshift from the peculiar velocity (Eq.~\ref{eq:zpec}).

\subsection{Zero-pointing TF and FP simultaneously}
Comparing Eqs.~(\ref{eq:dmu}) and (\ref{eq:eta}), it is apparent that
\begin{equation}\label{eq:Dmu-to-eta}
    \Delta\mu = -5\eta,
\end{equation}
which enables a simple transformation between the two (the ratio of luminosity distances is the same as the ratio of proper distances since the factors of $1+z_{\mathrm{obs}}$ cancel).

\begin{figure}
    \centering
    \includegraphics[width=\linewidth]{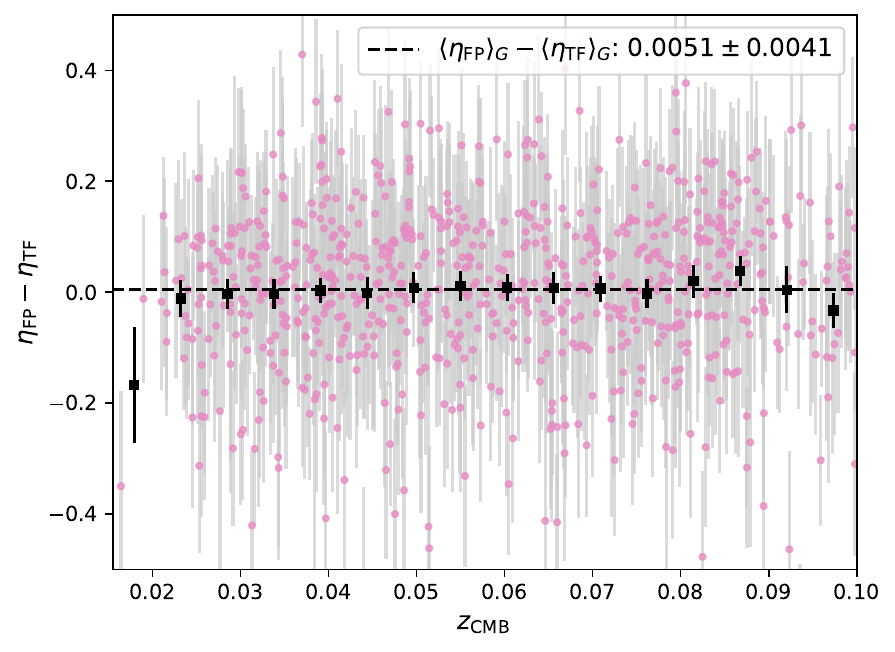}
    \caption{The difference between FP and TF $\eta$ in each group containing at least one of each type of galaxy. Individual $\eta$ differences are in pink, and the binned, weighted averages are in black. There is no significant trend with redshift. The black dashed line shows the average difference/offset.}
    \label{fig:FPshift}
\end{figure}

\begin{figure}
    \centering
    \includegraphics[width=\linewidth]{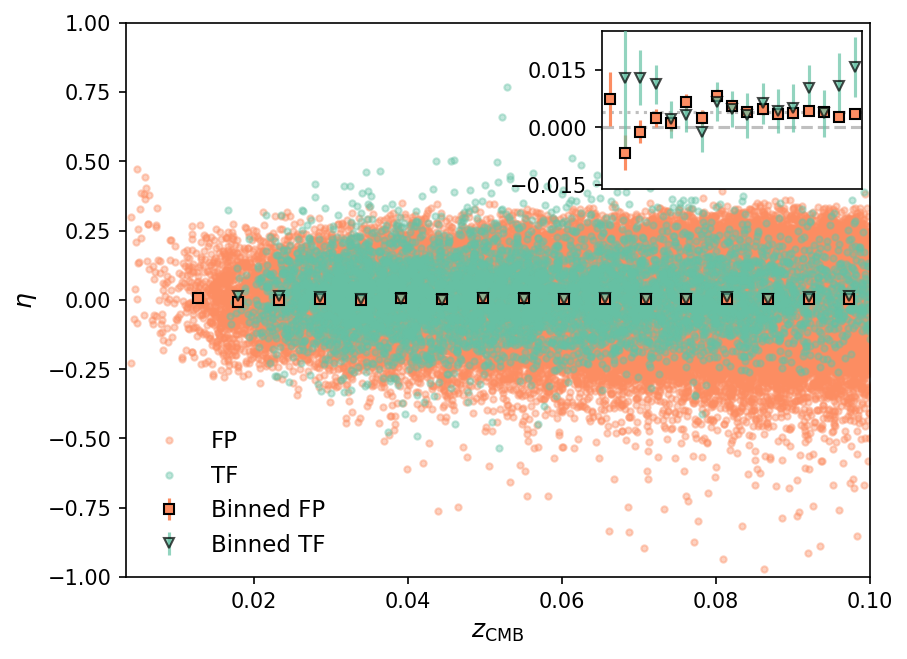}
    \caption{``Hubble diagram'' in $\eta$, which shows Hubble residuals as a function of redshift. We expect it to be flat in $z$, and centered on 0 when we use the same fiducial cosmology as was used to generate the catalogs. The inset axes zoom in to $\eta=0$ (gray dashed line); the mean (gray dotted line) is actually offset slightly, as already shown by the gray dotted line in Figure~\ref{fig:FP_group_corr}.}
    \label{fig:FP+TF_HD_corr}
\end{figure}

We must convert either TF to $\eta$ or FP to $\mu$ to be used together, and also with the calibrators we have access to in the DESI DR1 footprint (masers are physical size measurements, but all others are brightness measurements). 
Either conversion requires a cosmological model to convert $z$ to $D$, so a model must be assumed to calculate $\eta$ for FP, as well as to find $\mu_{\mathrm{model}}$ and therefore $\Delta\mu$ for TF.
For this model, we choose flat $\Lambda$CDM with the DESI fiducial cosmology of $\Omega_m=0.3151$ \citep[which comes from the Planck 2018 \texttt{base\_plikHM\_TTTEEE\_lowl\_lowE\_lensing} cosmology, meaning it was derived from the Planck 2018 baseline $\Lambda$CDM likelihood with the stated power spectrum data combination;][]{Planck2018}; the choice of $\Omega_m$ has no significant effect at such low redshift.
We arbitrarily set $H_0=100$ \kmsMpc{}, (\textit{i.e.}\ $h=1$); the value of $H_0$ is just a vertical shift to the Hubble diagram, so when we zero-point the data, the value we use here is absorbed into the zero-point.
Choosing a different $H_0$ would simply result in a different zero-point but the same measured $H_0$ post-calibration.

Both the TF and FP relations can be zero-pointed after fitting the respective relation, since it is just a shift to the respective Hubble diagram.
To fit each relation, an internal zero-point is still used but has no physical bearing until attached to independent distance measurements.
The TF and FP samples simply enforce no evolution in $\eta$ or $\Delta\mu$ across redshift bins. For a discussion of what this entails, see the FP \citep{DESI_DR1_FP} and TF \citep{DESI_DR1_TF} DR1 analyses.
We combine TF and FP measurements to find a global zero-point rather than treating them separately, but we can measure cosmology from each sample individually. 

\subsection{TF and FP alignment}\label{sec:TFFP_align}
Calculating the global zero-point is a two-step process.
Firstly, we carefully combine the TF and FP samples (denoting the combined sample ``PV''), ensuring the weighted average distances agree. Secondly, we calculate the difference between the weighted average of the combined sample and that of the calibrator sample.
This defines the global zero-point as 
\begin{equation}
    \eta_{\text{zp}} = \langle\eta_{\text{PV}}\rangle_{G'}-\langle\eta_{\text{cal.}}\rangle_{G'},
\end{equation}
where $G'$ is the subset of galaxies in groups containing both PV galaxies and calibrators, which are different to the groups containing at least one TF and FP galaxy, $G$. 
To convert the calibrators to $\eta$, we assume the same cosmological model as above.
We arbitrarily perform all calculations in $\eta$ and apply any zero-point offset to both $\mu$ and $\eta$ through the use of Eq.~(\ref{eq:Dmu-to-eta}).
We perform the analysis using the high-quality subsets of each sample (Sec~\ref{sec:cuts}), but the zero-point applies to all galaxies.

To align the two samples, we use the same definition, again taking advantage of galaxy groups, but the prescription of which sample to offset is arbitrary; either way, one of TF or FP is corrected to the other before both are corrected to the calibrator.
We choose to correct the TF sample to the FP sample, so the TF-FP offset is defined as
\begin{equation}
    \delta\eta= \langle\eta_{\text{FP}}\rangle_{G}-\langle\eta_{\text{TF}}\rangle_{G},
\end{equation}
and we transform $\eta_{\text{TF}}\rightarrow\eta_{\text{TF}}+\delta\eta$. 

\citet{DESI_DR1_TF} and \citet{DESI_DR1_FP} assumed the same fiducial cosmology as this work, so we expect $\delta\eta$ to be small, yet we still ensure there is no relative shift between the two Hubble diagrams.
In general, TF and FP samples are not guaranteed to be on the same internal zero-point or fiducial cosmology, and the peculiar velocities derived from a combined TF+FP sample are highly sensitive to the offset between them (an incorrect relative offset here would result in unphysical inflows/outflows that scale with redshift). 

Na\"{i}vely calculating the weighted average log-distance ratio or distance modulus of the FP or TF galaxies in groups results in a biased mean, as discussed in Sec.~\ref{sec:galaxy_groups} and shown clearly in Figure~\ref{fig:FP_group_corr}.
Thus, we first apply a correction ($\delta_{\text{FP}}$, defined in Eq.~\ref{eq:FP_mean_corr}, converted to $\eta$) to bring the mean of grouped FP galaxies back to the mean of all DESI DR1 FP galaxies through the transformation $\langle\eta_{\text{FP}}\rangle_G\rightarrow\langle\eta_{\text{FP}}\rangle_G-\delta_{\text{FP}}$ before calculating $\delta\eta$.
Note that the individual FP galaxies in groups are not corrected, as this step is just to calculate an unbiased offset between FP and TF.
The TF sample does not require a correction, as per Figure~\ref{fig:TF_group_corr}.
The offset was calculated using 784 groups (of size $\geq2$) that contained at least one FP and TF galaxy according to the expanded definition of \citet{Lim2017} groups.

The TF-FP offset is almost consistent with zero as expected, since both samples were generated with the same fiducial cosmology; see Figure~\ref{fig:FPshift}. 
In Figure~\ref{fig:FP+TF_HD_corr}, we show the Hubble diagram after shifting the TF sample by $\delta\eta$, to show the consistency of the two samples across redshift. 

We do not have a full covariance matrix for the DESI distances like we do with the Pantheon+ SNe. 
However, since we shift all TF distances by the same amount, we must account for the covariance this introduces. 
We achieve this by adding the variance of the correction, $\sigma^2_{\delta\eta}$, as a ``covariance block'', \textit{i.e.}\ all TF galaxies are slightly covariant with all others due to the coherent shift.
Therefore, we begin to construct a DESI PV covariance matrix that will eventually be used to calculate $H_0$, which at this point is entirely diagonal except for the TF block:\footnote{Since $\delta\eta$ was calculated using TF and FP galaxies in groups, there are extra covariance terms between TF and also FP galaxies in groups. However, these are small (especially when there are many galaxies in groups), and their effect is negligible; therefore, we consider only the first-order block covariance. The same is true for the global zero-point calculation, where we will also omit the second-order covariances.}
\begin{equation}\label{eq:cov_mat_TF}
    C_{ij} = \begin{cases} 
    \sigma^2_{\eta_i} + \sigma^2_{\delta\eta} & i = j\text{ and } i\in \text{TF}, \\
    \sigma^2_{\eta_i} & i = j\text{ and } i\notin \text{TF}, \\
    \sigma^2_{\delta\eta} & i \neq j\text{ and } i\land j\in \text{TF}, \\ 
    0 & \text{otherwise}.
    \end{cases}
\end{equation}
In order, the components of Eq.~(\ref{eq:cov_mat_TF}) are the diagonal inside the TF block, the diagonal outside the TF block, the off-diagonals inside the TF block, and finally the off-diagonals outside the TF block. Any operation after aligning the FP and TF samples, such as weighted averages, would use this covariance matrix rather than diagonal-only uncertainties, as we do in the next section.

\subsection{Calculation of the global zero-point from SNe Ia}
\begin{figure}
    \centering
    \includegraphics[width=\linewidth]{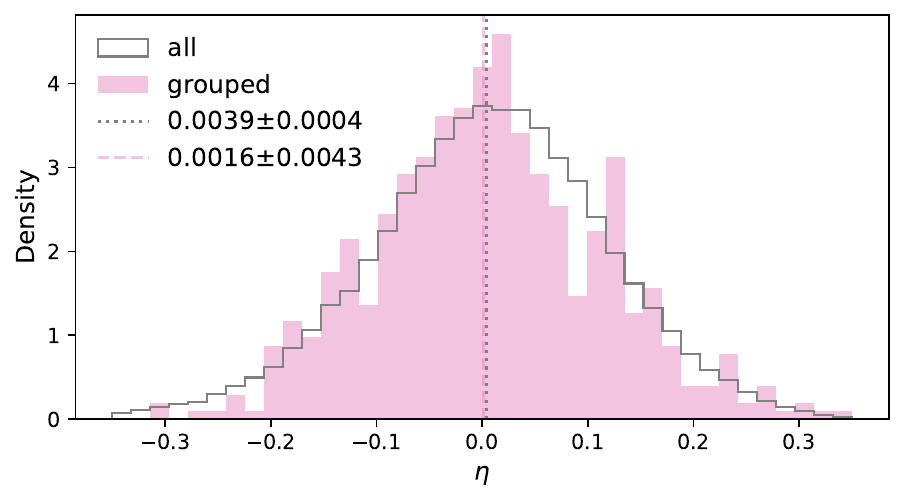}
    \caption{Comparison of the distributions of all DESI PV $\eta$ (gray) and of the galaxies in groups with SNe (pink). The means (dashed lines) disagree by $<1\sigma$, but the K--S test still shows some evidence that they are different distributions.}
    \label{fig:groupcorr}
\end{figure}

The second step is to define the global zero-point using the combined, consistent TF+FP samples.
We start by ensuring the population of groups that host PV galaxies, along with SNe, is the same as the full population.
Unlike the case of all groups containing FP galaxies versus all FP galaxies, our results on the difference for the groups containing SNe versus all galaxies are less clear (Figure~\ref{fig:groupcorr}).
Specifically, the likelihood that the populations are different \textit{is} high, with a $p$-value of 0.05 from the weighted K--S test, but the weighted mean difference, $\delta_{\text{SN}}=\langle\eta_{\text{PV}}\rangle_{G'}-\langle\eta_{\text{PV}}\rangle$, is still consistent with zero.
Since there is some evidence for a difference in population, despite the consistency with zero, we applied the group correction in a similar fashion to FP (via $\langle\eta_{\text{PV}}\rangle_{G'}\rightarrow\langle\eta_{\text{PV}}\rangle_{G'}-\delta_{\text{SN}}$).
There is a hint that it may become more necessary or of higher significance with a larger PV and calibrator overlap, using the current grouping method.
If the source of the discrepancy between grouped and non-grouped is indeed the differing completeness of the group catalog and DESI observations, then we expect this step will not be necessary once a DESI-based group catalog is available.

\begin{figure}[t!]
    \centering
    \includegraphics[width=\linewidth]{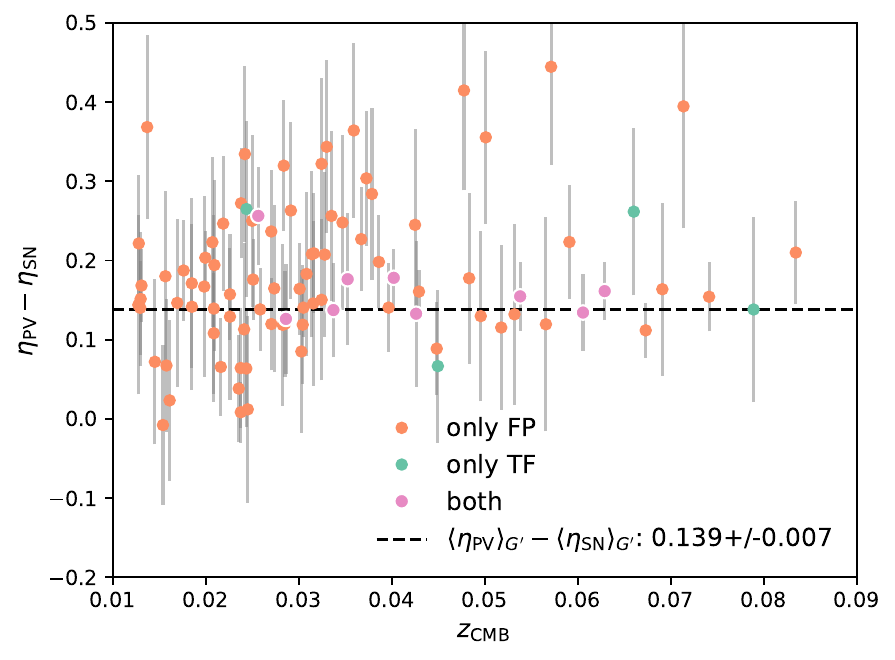}
    \caption{The difference between the log-distance ratio given by DESI and Pantheon+ SNe Ia for individual groups (for visualization only). The average difference between the two is the zero-point (black dashed line). We shift the DESI PV sample to match the SN.}
    \label{fig:SN_zp}
\end{figure}

\begin{figure}[t!]
    \centering
    \includegraphics[width=\linewidth]{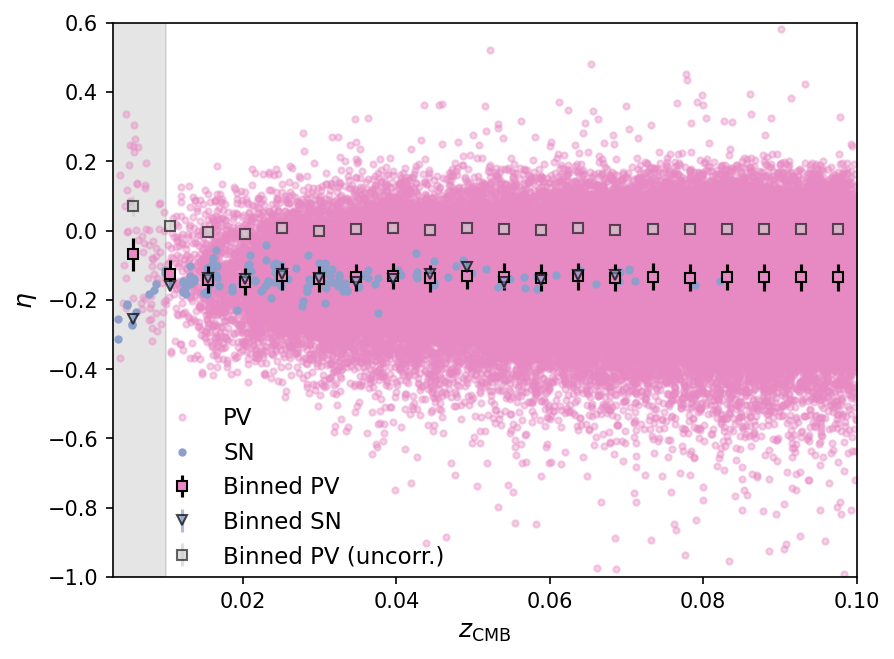}
    \caption{Combined TF+FP (PV) log-distance ratio Hubble diagram with Pantheon+ SNe Ia. The DESI PV log-distance ratios (pink circles, squares) have been shifted to the SN Ia (blue circles, triangles) zero-point. The binned, uncorrected DESI $\eta$ (gray squares) are displayed for comparison. The shaded region ($z<0.01$) was excluded from the zero-point fit because the calibrators and DESI sample show disagreement.}
    \label{fig:SN_PV_HD}
\end{figure}

The global zero-point is defined as the difference in weighted averages of the combined DESI PV sample and the SN Ia sample, respecting full covariance, including subtle TF covariance from Sec.~\ref{sec:TFFP_align}.
Rather than the scalar weighted average (equivalent to a diagonal covariance matrix), in general, $\langle x\rangle=\sigma^2_{\langle x\rangle}(\textbf{1}^TC^{-1}\boldsymbol{x})$ where $C$ is the covariance matrix, \textbf{1} a vector of ones, and $\sigma^2_{\langle x\rangle}=(\textbf{1}^TC^{-1}\textbf{1})$ is the variance of the weighted mean. 
This results in a zero-point of $\eta_{\text{zp}}=0.139\pm0.007$.
Since we shift all DESI galaxies by this zero-point, this again introduces covariance that we account for by adding the variance, $\sigma^2_{\eta_{\text{zp}}}$ as a block to the DESI covariance matrix.
This time, the block is the same size as the full matrix and acts to propagate the uncertainty from the entire SN distance ladder as an irreducible uncertainty on our own $H_0$ measurement.
Thus, in total, the DESI covariance matrix is
\begin{equation}\label{eq:cov_mat}
    C_{ij} = \begin{cases} 
    \sigma^2_{\eta_i} + \sigma^2_{\eta_{\text{zp}}} + \sigma^2_{\delta\eta} & i = j\text{ and } i\in \text{TF},\\
    \sigma^2_{\eta_i} + \sigma^2_{\eta_{\text{zp}}} & i = j\text{ and } i\notin \text{TF}, \\
    \sigma^2_{\eta_{\text{zp}}} + \sigma^2_{\delta\eta} & i \neq j\text{ and } i\land j\in \text{TF}, \\ 
    \sigma^2_{\eta_{\text{zp}}} & \text{otherwise},
    \end{cases}
\end{equation}
which is the same as Eq.~(\ref{eq:cov_mat_TF}) with $C_{ij}+\sigma^2_{\eta_{\text{zp}}}$. Any operation on the calibrated sample, such as estimating $H_0$, should now include this covariance matrix.

In Figure~\ref{fig:SN_zp}, we show the individual offset for each group, this time differentiating between which DESI PV type was used.
This exercise is purely for visualizing any possible redshift trends, as the individual differences do not account for covariance or weights which is why the weighted average appears small given the relatively large positive scatter.
There is no significant evolution in the differences beyond $z>0.01$.

Figure~\ref{fig:SN_PV_HD} shows the zero-pointed Hubble diagram in $\eta$-space of the DR1 PV sample along with the SNe Ia used to zero-point it.
When aligning the TF and FP samples, we performed a redshift cut of $0.0033<z<0.1$ consistent with the other DR1 analyses (which had no bearing on the alignment).
When we zero-pointed the combined sample, we increased the lower redshift cut to $z>0.01$, consistent with the DR1 clustering analyses of \citet{DESI_DR1_PV_dens_vel_corr}, \citet{DESI_DR1_PV_max_like} and \citet{DESI_DR1_PV_power_spec} --- the region that was cut is shaded gray in Figure~\ref{fig:SN_PV_HD}.
Below this lower redshift cut, there appears to be a disagreement between DESI PV sample distances and those from calibrators.
We expect a low-redshift evolution of Hubble residuals due to local large-scale structure \citep[according to velocity-field reconstructions, as discussed in][]{Said2025} and the volume scattering effect \citep[SNe more likely to scatter from larger volumes to smaller due to PVs than vice versa;][]{Pplus_cosmo}.
These effects cause the downward turn in the SN log-distance ratios at low redshift.
However, the scatter seen in the DESI PV sample is in the opposite direction and adds a bias when relying on calibrators in that redshift region.
Allowing these very low redshifts for the SN sample has an appreciable effect (see Sec.~\ref{sec:systematics}), despite most of the power coming from larger redshifts.
Additionally, for the other calibrators, which mostly or wholly occupy the biased region, the zero-point is strongly biased and cannot be used.
Thus, we apply a simple redshift cut to all zero-point measurements.

\subsection{Calculation of the global zero-point from alternative calibrators}
SNe Ia are the natural choice of fiducial calibrator because they are numerous and cover the entire redshift range of the DESI PV sample.
However, SNe Ia are just one choice.
Within the DESI DR1 PV sample, we also found either direct galaxy or group overlap with Cepheid, maser and SBF distance measurements.
For the FP EDR release, \citet{Said2025} used a single calibration based on the distance to the Coma cluster that we also used for a direct comparison/update.
The measurement that was used was SBF-based, from \citet{Jensen2021}; however, there has since been an updated measurement by \citet{Scolnic2025} using SNe Ia with greatly improved precision and acting as a parallel to our Pantheon+ SN-based fiducial measurement.

\begin{figure}
    \centering
    \includegraphics[width=\linewidth]{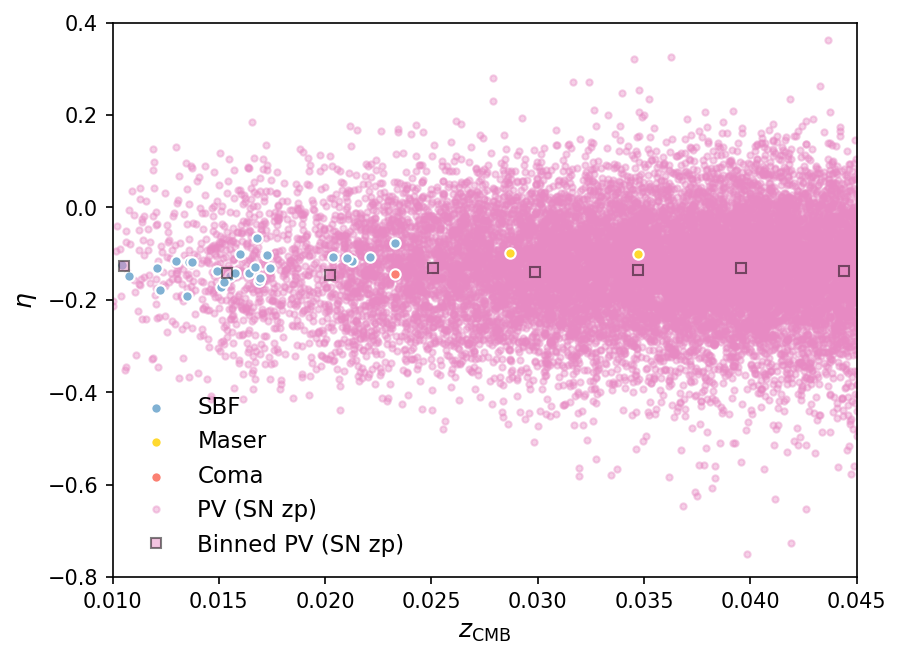}
    \caption{Alternative calibrators compared to the DESI log-distance ratio Hubble diagram. The white-outlined circles represent each calibrator, and the DESI PV values (pink circles, squares) are displayed on the SN zero-point for comparison.}
    \label{fig:calHD}
\end{figure}

Again, we performed the same global zero-point process for each calibrator. 
For most cases, there was insufficient data to conclude whether the grouped galaxies containing a calibrator required correction back to the population mean.
Figure~\ref{fig:calHD} shows each of the alternative calibrators in comparison to the DESI sample.
Due to the $z>0.01$ cut, no Cepheids remained in the alternative calibrator sample, as the four candidates were all below $z=0.01$.
Table~\ref{tab:cal_zps} shows the values of the individual zero-points that were possible to measure.
The agreement between SBF and fiducial SN Ia measurements is notable; the fiducial zero-point of $\eta_{\text{zp}}=0.139\pm0.007$ and the SBF measurement of $\eta_{\text{zp}}=0.134\pm0.008$ disagree by only 0.005, or 25 mmag in distance modulus, and have similar statistical uncertainty despite there being a quarter as many groups with SBF (\nSBF{}) as SNe Ia (\nSN{}).

\begin{deluxetable*}{cCCCC}
\tablewidth{0pt}
\tablehead{
\colhead{Calibrator} & \colhead{$N_{\mathrm{cal}}$} & \colhead{$\eta_{\text{zp}}$ } & \colhead{$\sigma_{\eta_{\text{zp}}}$ (stat.)}
 }
\tablecaption{Value of the zero-point for each calibrator.\label{tab:cal_zps}}
\startdata
SNe Ia & \nSN\ & 0.139 & 0.007 \\
SBF & \nSBF\ & 0.134 & 0.008\\
SNe Ia \& SBF & \nSNSBF\ & 0.136 & 0.006\\
Masers & 2 & 0.139 & 0.033\\
Coma (SNe Ia) & 1 & 0.116 & 0.012\\
Coma (SBF) & 1 & 0.119 & 0.027 \\
\enddata
\end{deluxetable*}

It is apparent that the SN Ia and SBF zero-points are in very good agreement, and are both calibrated by Cepheids \citep{Jensen2021, Tully2023}, so the combination should provide the tightest statistical constraints possible for DR1 (while still sharing the systematic errors due to the shared Cepheid-based calibration).
We do not have an estimate of covariance for any other calibrator apart from SNe, so for the combination with SBF, we create a new SN+SBF covariance matrix containing the original SN matrix and an additional diagonal SBF block.
As we discuss in Sec.~\ref{sec:cosmology}, the lack of covariance means that uncertainties are underestimated compared to a full, non-diagonal treatment.
Accounting for the galaxy groups that share SBF and SN galaxies, there are \nSNSBF\ groups we can use to calibrate the zero-point, a 10\% increase to the SN Ia only sample size.
We still detect no evidence of a biased group-versus-full population with the slightly increased sample size.
The combined SN+SBF zero-point shifts to $0.136\pm0.006$. 

\section{Cosmological results}\label{sec:cosmology}
Once the combined catalog of TF and FP has been zero-pointed, we fit the standard distance modulus Hubble diagram to constrain $H_0$. 
$H_0$ is usually fit in the redshift range of $0.0233<z<0.15$, but we are limited to an upper bound of $z=0.1$, which we do not expect to have a major effect.
The lower limit of $z=0.0233$ was originally set to reduce the impact of PVs on the measurement, and above this cut is thus deemed the ``Hubble flow'' \citep{Riess2016}.
We also investigate the effects of differing lower limits in Sec.~\ref{sec:systematics}.

We transform our redshifts to the ``Hubble diagram'' frame $z_{\text{HD}}$ by correcting the CMB-frame redshifts for the peculiar velocities of the DESI PV galaxies as given by velocity-field reconstruction.
The code we use, \texttt{pvhub},\footnote{\url{https://github.com/KSaid-1/pvhub}.} is based on the 2M++\footnote{The 2M++ reconstruction \citep{Carrick2015} is based on the 2M++ redshift survey data compilation \citep{Lavaux2011}.  2M$++$ includes data from 2MRS, 6dFGS, and SDSS---similar to the \citet{Lim2017} catalogs---and reconstructs large-scale coherent motions out to a radius of 200 \hMpc{}.} velocity-field reconstruction \citep{Carrick2015} with some modifications as described in \citet{Carr2022}.
This is an independent method to the PVs we aim to measure from the DESI sample, and is just for the $H_0$ fit, so there is no inconsistency or double-counting in this approach.
Correcting the redshifts for large-scale coherent motion (the reconstruction necessarily smooths on the Mpc scale, so information smaller than that scale is not corrected) is the best practice for fitting $H_0$, as shown by \citet{Peterson2022}.
In fact, velocity-field reconstruction can mitigate the effects of PVs at low redshift, the reason for which we cut $z<0.0233$ from the SN Hubble diagram.
$H_0$ has been constrained from a two-rung distance ladder with Cepheid variable stars themselves --- without the need for SNe Ia --- in \citet{Kenworthy2022} and recently to impressive precision using a new, state-of-the-art constrained realization simulation rather than linear theory in \citet{Stiskalek2025}.

For the $H_0$ fit, we use $\mu$ rather than the log-distance ratios used to calculate the zero-points.
We then compare the set of measured $\mu$ to the cosmological expectation, $\mu_{\mathrm{model}}$, as given by a Taylor series expansion of the luminosity distance in a spatially flat Universe, 
\begin{equation}
\begin{array}{l}
    \mu_{\mathrm{model}} = 25+5\displaystyle\mathrm{log}_{10}\left\{\frac{c z_{\mathrm{HD}}}{H_0}\right.\\[0.5em]
    \qquad\displaystyle\left.\times \left[1 + \frac{z_{\mathrm{HD}}}{2} (1 - q_0) - \frac{z^2_{\mathrm{HD}}}{6} (1 - q_0 - 3q^2_0 + j_0) \right]\right\}
\end{array}
\end{equation}
where $q_0$ and $j_0$ are the deceleration and jerk parameters, respectively. 
Usually, analyses fix $q_0=-0.55$ and $j_0=1.0$, which is consistent with a Universe where $(\Omega_m,\Omega_\Lambda)=(0.3,0.7)$, such as in \cite{Said2025} and \cite{SH0ES22}.
Since our low-redshift data has little constraining power on these parameters, especially $j_0$, for most of our fits we fix $q_0=-0.527$ to be consistent with the fiducial cosmology of $\Omega_m = 0.3151$, and $j_0=1$.
We also allow $q_0$ to be a free parameter, but we reiterate that there is very little constraining power in the redshift range of the DESI PV Survey.
When \citet{SH0ES22} allowed $q_0$ to vary, and with the inclusion of high-redshift SNe Ia, they found no effect on $H_0$ and that $q_0=-0.510\pm0.024$.

We perform the fitting of our Hubble diagram using the static nested sampling procedure \texttt{dynesty} \citep{Speagle2020, Koposov2024, Skilling2004, Skilling2006} via the Bayesian inference Python package \texttt{bilby} \citep{Ashton2019}. We initialize the sampler with a uniform prior on $H_0$ of [50, 90] \kmsMpc. 
The likelihood function we maximize is
\begin{equation}\label{eq:likelihood}
    \mathcal{L} = (2\pi)^{-N/2}|C|^{-1/2}\exp\left(-\frac{1}{2}\boldsymbol{\Delta\mu}^TC^{-1}\boldsymbol{\Delta\mu}\right),
\end{equation}
where $N$ is the number of PV galaxies, and the covariance matrix, defined in Eq.~(\ref{eq:cov_mat}), has been converted to $\mu$.
We also modify the covariance by adding another block representing the $H_0$ systematic uncertainty (discussed in detail in the following section) converted to magnitudes. 
The $\mu$ measurement uncertainties are large compared to the uncertainties on the spectroscopic DESI redshifts, which justifies our choice to ignore redshift uncertainties in our fits.
Since this covariance matrix is large, we utilize the Woodbury Identity to decompose the covariance matrix into smaller, more manageable matrices. Refer to Sec.~2.7.3 of \citet{Woodbury} for a complete description of this technique.

\subsection{Tests of systematics}\label{sec:systematics}
\begin{figure}[t!]
    \centering
    \includegraphics[width=0.65\linewidth]{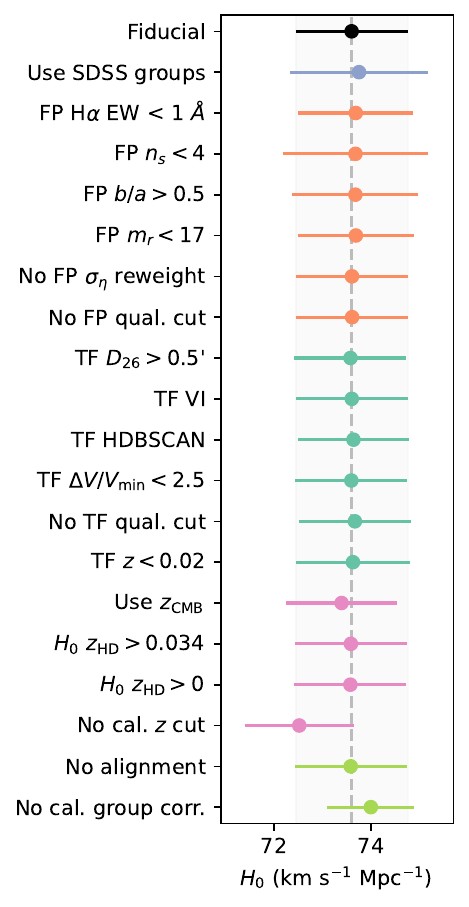}
    \caption{Whisker plot of all systematic tests. Each point corresponds to a single change to the fiducial zero-pointing and $H_0$ fitting pipeline. The black point, gray-shaded region and gray dashed line all represent the fiducial measurement, and each color corresponds to the systematic group described in Sections~\ref{subsec:syst_group} to \ref{subsec:syst_zp-ing}. The weighted standard deviation of all tests results in $\sigma_{H_0}\;\text{(syst.)}=\HsystSN{}$ \kmsMpc{}.}
    \label{fig:syst_whisker}
\end{figure}

In this section, we describe our investigation into potential systematics.
In what follows, we give a brief description of the tests and then estimate the final systematic uncertainty, only for the fiducial SN-only calibration.
In most cases, where necessary, we run the full analysis pipeline since some of the changes affect which galaxies are grouped.
Figure~\ref{fig:syst_whisker} shows the impact of each systematic in the order they are introduced below (each color represents a different subsection).

\subsubsection{Grouping algorithm}\label{subsec:syst_group}
One of the largest potential systematics comes from the assignment of DESI galaxies to groups, which was done primarily to increase the calibrator pool.
The modified algorithm we use, based on the \citet{Lim2017} catalogs, can be modified to include more or fewer iterations, but a more telling test is to use an original catalog instead.
In the DESI footprint, only the SDSS catalog is deep and wide enough to be used. 
Therefore, for this test, we replaced the custom grouping method with the unmodified SDSS group catalog.

\subsubsection{FP catalog systematics}
\citet{DESI_DR1_FP} performed many internal validation tests by propagating different data quality/selection cuts and parameter choices on the recovered FP relation.
We propagate the effects of those tests with the largest impact on the recovered FP parameters as an upper-bound on the FP-catalog-based systematic uncertainty.
All of these systematics relate to the different methods of eliminating spiral or non-elliptical galaxies from the sample.
Specifically, these are requiring the equivalent width (EW) of the H$\alpha$ emission line to be $<1$ \AA; the S\'{e}rsic index $n_s<4$; the axial ratio $b/a>0.5$; and $r$-band magnitude $m_r<17$ mag.  

Also in this category are the improvements made to FP since EDR: the group correction and Gaussianization of the log-distance ratio uncertainties $\sigma_\eta$.
As described in \citet{DESI_DR1_FP}, the detection of a correlation between $\eta$ and group richness is very strong, so this correction is essential and we do not consider it as a systematic.
However, we do remove the Gaussianization of the log-distance ratio uncertainties as a systematic test.

Finally, we relax the quality cut on FP galaxies, meaning we use the sample of FP galaxies without the iterative outlier rejection procedure applied.

\subsubsection{TF catalog systematics}
\citet{DESI_DR1_TF} made several analysis choices, such as different criteria for rotation velocity repeatability and spiral classification.
Similar to FP, the TF catalog is regenerated assuming different cuts, where we exclude galaxies smaller than $R_{26}=0.5^\prime$; alter the spiral classification to visual inspection only; increase the requirement on rotation velocity repeatability to $\Delta V/V_{\text{min}}<2.5$; use the Hierarchical Density-Based Spatial Clustering of Applications with Noise (HDBSCAN) algorithm for TF outlier rejection (or, rather, finds the data points most likely belonging to the TF relation based on their TF parameter space clustering).

Also in this category is the reduction of the quality cut to only exclude dwarf galaxies, rather than using the default rotation velocity selection.

Finally, we consider restricting the upper redshift limit of the TF sample to $z<0.05$, consistent with the DR1 clustering analyses. 

\subsubsection{Redshifts}
Unlike the method used in \citet{Said2025}, we use the peculiar-velocity-corrected $z_{\text{HD}}$ to measure $H_0$ in the fiducial case, so the systematic test is to undo this and use $z_{\text{CMB}}$ instead.
This not only alters which galaxies pass the redshift cut, but also coherently shifts redshifts due to the effects of local large-scale structure.

We also consider both the redshift range for measuring $H_0$ and for the calibrators we use.
Following \citet{Said2025}, we perform no lower cut to the $H_0$ fit, $z_{\text{HD}}>0$ and a cut of $z_{\text{HD}}>0.034$ for comparison.

Finally, we consider the case where we undo the $z>0.01$ cut to the calibrators (the shaded region in Figure~\ref{fig:SN_PV_HD}). 
Allowing the zero-point to be calculated using this region is the largest contributor to the systematic uncertainty in the DR1 $H_0$ measurement, although still less than $1\sigma$ from the fiducial case.

\subsubsection{Zero-pointing process}\label{subsec:syst_zp-ing}
The final tests we perform relate to the zero-pointing process as described in Sec.~\ref{sec:zp_process}.
We start by aligning the FP and TF samples because there is no guarantee that the weighted means of each sample agree initially. 
However, we find that the correction is negligible, so one test is not to perform this alignment.
While $H_0$ is insensitive to this test, the peculiar velocities are more sensitive (which would propagate to the clustering analyses).

The other main correction we performed was to correct the weighted mean of the DESI PV galaxies in groups that contain SNe Ia back to the population weighted mean ($\delta_{\text{SN}}$). 
Since the $p$-value from the weighted K--S test showed marginal evidence of the difference in distributions, we also test the case of not performing the correction. 
When performing other systematic tests, $\delta_{\text{SN}}$ was sometimes non-negligible (larger value and/or lower $p$-value), implying stronger evidence of a difference in grouped and non-grouped populations of galaxies.
Using a different set of galaxy groups results in a different correction because a) the mean log-distance ratio is different, and b) there are not many groups containing calibrators to begin with.
The set of galaxy groups naturally changes with different data cuts or other changes earlier in the pipeline.
However, even when $\delta_{\text{SN}}$ varies, the systematic tests show very little variation in $H_0$.
It is because we apply the group correction that the $H_0$ variation is so small. 
After all, we are always correcting for any bias introduced by relying on grouped galaxies, even though in the fiducial case it appeared unnecessary.
Without the group correction, the zero-point is biased by a small amount in the fiducial case, but by a large amount in other cases, so in retrospect, we consider it essential.
This is the second largest systematic.

\subsubsection{Total systematic uncertainty}
From all of these tests, shown in Figure~\ref{fig:syst_whisker}, we find a weighted standard deviation from the fiducial analysis of \HsystSN\ \kmsMpc{}. 
This represents our estimate of the systematic uncertainty from the FP+TF measurements and our zero-point methodology.
This is a slight decrease compared to the EDR FP finding of 0.49 \kmsMpc{} despite the addition of TF.
This is unsurprising given the more careful treatment of FP, which dominates the DR1 measurement by 10:1 FP to TF galaxies.

However, our fiducial measurement uses SH0ES/Pantheon+ SNe Ia, so our measurement also inherits the systematics from the three-rung distance ladder.
When we treat the full statistical+systematic uncertainties in the SH0ES/Pantheon+ covariance matrix as an irreducible uncertainty floor, our estimate of systematic uncertainty is in total \HtotsystSN{} \kmsMpc{}.

\subsection{Choice of calibrator}
The variation of $H_0$ due to the choice of calibrator is shown in Figure~\ref{fig:H0_whisker}. 
Each calibrator (except masers) is itself calibrated with Cepheid distances in some way; the SBF distances are calibrated using both Cepheids and TRGB \citep{Jensen2021}, and the two distances to Coma were measured using SBF and SNe Ia.
This is also consistent with the masers, which are used to calibrate the Cepheids.
The dependence on Cepheids and masers is just a quirk of the DR1 PV sample, and in future releases, we hope to make use of different calibrators.

Similar to the fiducial case, the other calibrators we use also have systematic uncertainties.
We derive the systematic floor for the other calibrators in the same way as for the Pantheon+ SNe Ia, \textit{i.e.}\ we simply use diagonal calibrator matrices.
However, in the absence of a full covariance matrix for each, these uncertainty floors are likely underestimated.
Some of the increase in precision in the SN+SBF measurement compared to the SN-only measurement is real, since there are more calibrators, but it is still artificially small due to the lack of SBF covariance.
Therefore, we continue to quote the SH0ES/Pantheon+ SN measurement as the best measurement of this work since it has the most accurate uncertainty.

\subsection{\texorpdfstring{$H_0$}{H0} results with fixed \texorpdfstring{$q_0$}{q0}}
\begin{figure*}[ht!]
    \centering
    \includegraphics[width=0.6\linewidth]{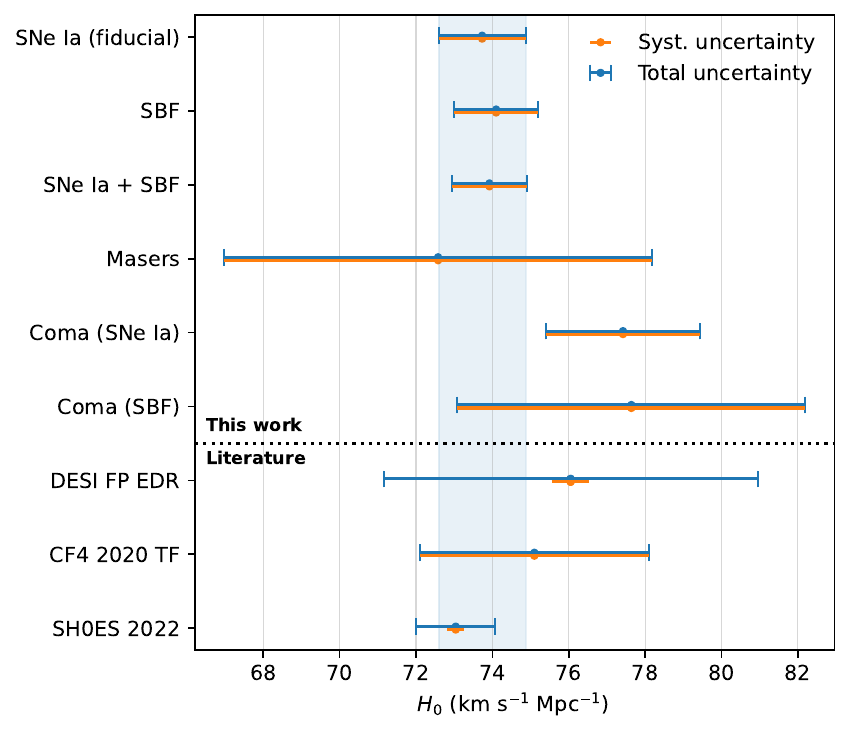}
    \caption{Comparison of $H_0$ constraints from each choice of calibrator and from the literature. Each DESI DR1 measurement comes from the combined FP and TF relations calibrated by each of the calibrators shown above the dotted line. For each measurement, we show the total uncertainty (blue) and highlight the systematic component (orange). The blue shaded region corresponds to the fiducial measurement for comparison.
    }
    \label{fig:H0_whisker}
\end{figure*}

\begin{figure*}[ht!]
    \centering
    \includegraphics[width=0.7\linewidth]{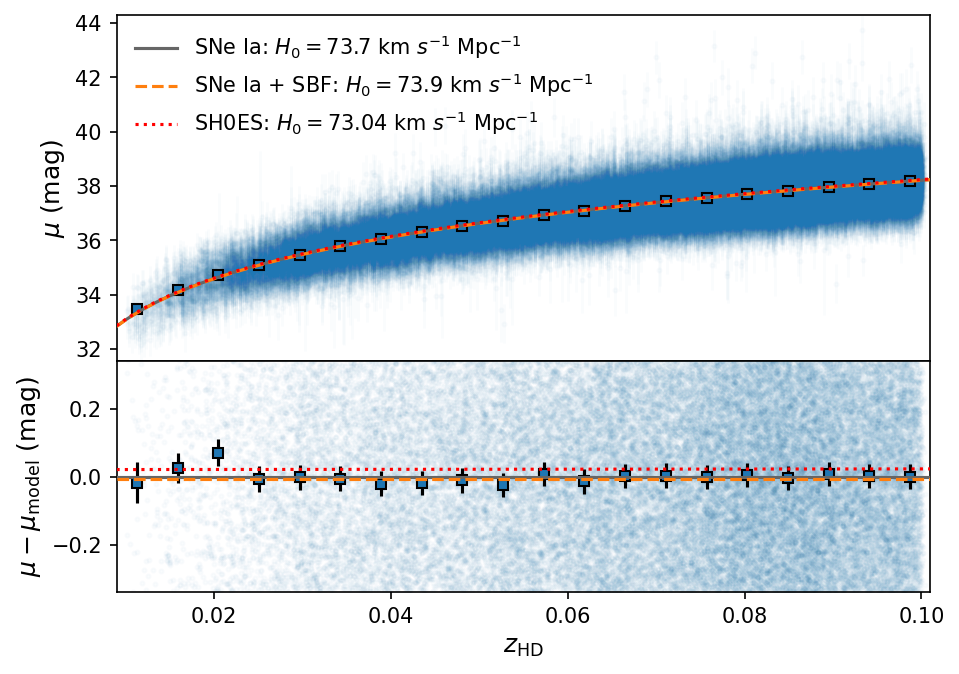}
    \caption{Hubble diagram displaying the DESI distance moduli (faint blue points) with our fiducial best fit using the SN Ia zero-point with fixed $q_0$ (gray line), with the SN+SBF and SH0ES results also shown for comparison (orange dashed and red dotted line, respectively). The binned data (blue squares) were calculated from the DESI covariance matrix built to fit $H_0$. }
    \label{fig:HD_H0fits}
\end{figure*}

For each of our measurements, the total reported uncertainty constitutes a combination of the statistical uncertainty from our fit to the Hubble diagram, along with systematic uncertainties. We account for three sources of systematic uncertainty: 1) Uncertainty on the calibration of the zero-point added to the Hubble diagram distance moduli, $\mu$. 2) A TF-FP alignment (as described in Sec.~\ref{sec:TFFP_align}) zero-point uncertainty on only TF distance moduli. 3) The total uncertainty found from our systematic tests (see Sec ~\ref{sec:systematics}). We note that the first two systematic uncertainties ought to be treated as a systematic floor uncertainty that cannot be outperformed even with the addition of more distance moduli measurements. For this reason, our $H_0$ constraints are never tighter than those achieved from the calibrator by itself.

For each calibrator type, we provide an approximate breakdown with the statistical uncertainty, statistical plus zero-pointing uncertainty, and the total uncertainty. The uncertainty in each of these cases is found by setting all uncertainties --- aside from the source(s) of uncertainty being measured --- to zero. These results are displayed in Table~\ref{tab:H0summary}. Due to this method of isolating different components and the stochasticity of the nested sampling, there are slight variations each time we run the fits. This results in slight variations to both the statistical uncertainty (which should be constant for each fit) and the value of $H_0$, which is why this is an approximate error budget. These variations are negligible compared to the total uncertainties.

Our fiducial result, from the SNe Ia only zero-point, is $H_0 = \HSN\pm\HstatSN\;\text{(stat.)}\pm\HtotsystSN\;\text{(syst.)}=\HSN\pm \HtotSN{}$ \kmsMpc. Figure~\ref{fig:HD_H0fits} shows the resulting Hubble diagram, along with the SNe Ia + SBF result, and Figure~\ref{fig:H0_whisker} shows results derived using the full set of calibrators listed in Sec.~\ref{sec:calibrators}, along with other $H_0$ measurements from previous studies. 

The previous studies to which we compare (also shown in Figure~\ref{fig:H0_whisker}) are the FP EDR measurement using 4,191 FP galaxies calibrated using the SBF distance to Coma $H_0 = 76.05\pm0.35\;\text{(stat.)}\pm0.49\;\text{(syst.)}\pm4.86\;\text{(cal. stat.)}$ \kmsMpc{}, the CosmicFlows-4 measurement that used $\sim$10,000 TF galaxies $H_0 = 75.1\pm0.2\;\text{(stat.)}\pm3\;\text{(syst.)}$ \kmsMpc\ \citep{Kourkchi2020}, and the SH0ES/Pantheon+ measurement $H_0 = 73.04\pm1.04$ \kmsMpc\ that we expected to recover in the fiducial case. 
The EDR FP analysis treated the uncertainty due to calibration as statistical, whereas the CosmicFlows-4 analysis treated it as systematic. Our approach of treating the zero-point uncertainty (which depends on the number and precision of the calibrators) as irreducible is similar to the latter.

\begin{figure}[tb]
    \centering
    \includegraphics[width=0.45\textwidth]{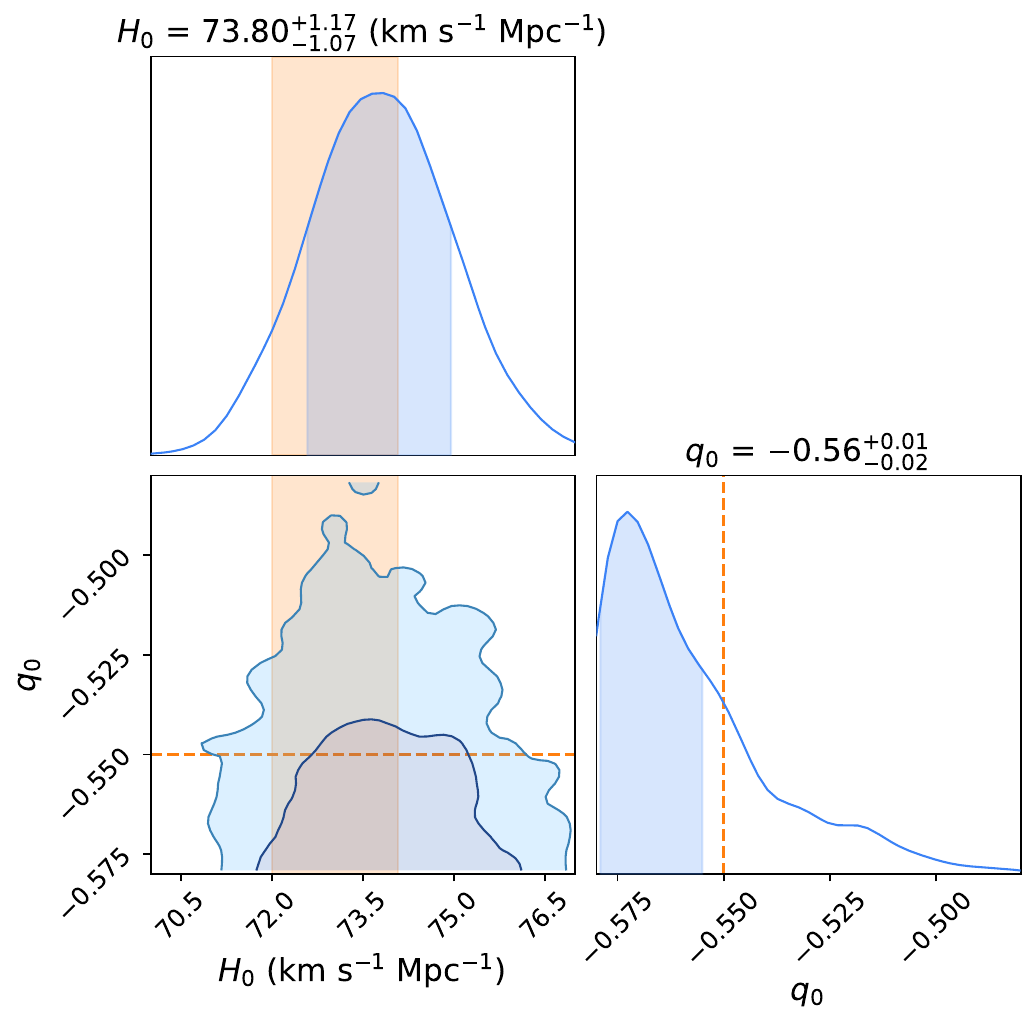}
    \caption{Posteriors on cosmological parameters from the DESI PV sample zero-pointed to SNe Ia (blue) compared with the fiducial SH0ES result (orange) for which $q_0$ was fixed to $-0.55$.}
    \label{fig:H0q0_corner}
\end{figure}

\begin{deluxetable*}{cCCCCC}
\tablewidth{0pt}
\tablehead{
\colhead{Calibrator} & \colhead{$N_{\mathrm{cal}}$} & \colhead{$H_0$} & \colhead{$\sigma_{H_0}$ (stat.)} & \colhead{$\sigma_{H_0}$ (zp)} &  \colhead{$\sigma_{H_0}$ (total)} \\
 &  & \kmsMpc\ & \kmsMpc\ & \kmsMpc\ & \kmsMpc\
 }
\tablecaption{Summary of $H_0$ constraints and approximate error budget for each calibrator.\label{tab:H0summary}}
\startdata
SNe Ia        & \nSN\ & \HSN\ & \HstatSN\  & \HtotsystSN\ & \HtotSN\ \\[0.3em]
SBF           & \nSBF\ & \HSBF\ & \HstatSBF\ & 1.1 & 1.1 \\[0.3em]
SNe Ia \& SBF & \nSNSBF\ & \HSNSBF\ & \HstatSNSBF\ & 0.98 & 1.0 \\[0.3em]
Masers        & 2 & 72.6 & 0.06 & 5.6 & 5.6 \\[0.3em]
Coma (SNe Ia) & 1 & 77.4 & 0.07 & 2.0 & 2.0 \\[0.3em]
Coma (SBF)    & 1 & 77.6 & 0.06 & 4.6 & 4.6 \\[0.3em]
\enddata
\tablecomments{The zero-point uncertainty, $\sigma_{H_0}$ (zp), is the uncertainty on $H_0$ arising from the irreducible zero-point uncertainty. The total uncertainty additionally includes the estimated systematic uncertainty on the fiducial measurement, \HsystSN{} \kmsMpc{}, which we assume is the same for each calibrator, and a contribution from the TF-FP alignment (smaller than the statistical uncertainty).}
\end{deluxetable*}

In general, the results are as expected.
Due to the factor of 25 more galaxies in the DESI DR1 PV sample (104,000) compared to the EDR FP sample (4,191), the statistical uncertainty has drastically reduced from 0.35 \kmsMpc{} to \HstatSN{} \kmsMpc{}, almost exactly by the factor of $1/\sqrt{N}$ expectation.
This is excluding the ``calibration statistical'' uncertainty from the EDR, which we instead treat as a systematic uncertainty. 
When we use the same calibrator as the EDR analysis, we observe a value consistent with but slightly larger than the EDR analysis and the SBF themselves, but the SN-calibrated Coma distance does not agree with the SN-only value.
Ideally, we should be calibrating the PV relations on as many clusters as possible, rather than one (or two) precisely measured cluster(s), so our main results come from the SNe and SBF.
The consistency of the SBF and SN Ia calibration of the distance to Coma shows that the relatively high $H_0$ may be a property of calibrating from just one cluster rather than a range of positions and redshifts.
When using the two masers that overlap with DR1, we see a lower $H_0$ than the SNe Ia that were calibrated using masers, and the large uncertainty from relying on only two masers is apparent.
The SNe Ia and SBF were all calibrated from Cepheids (which were calibrated using masers) and all show good consistency, as expected.
The combination of SNe Ia and SBF that is allowed from being on the same distance ladder shows an increase in precision due to the SBF adding statistical power.

The central values of our fiducial measurement and SH0ES are reasonably consistent, although our result is slightly higher.
All of the SH0ES/Pantheon+ systematics (for the SN subset we use here) have been incorporated into our analysis through the SN covariance matrix, which is why our systematic uncertainty is approximately the same size as their total uncertainty.
The use of over 100,000 DESI PV tracers  (despite their large intrinsic scatter), and a measurement that is robust to systematics (our tests revealed $\lesssim1\sigma$ shifts to $H_0$), allows us to improve on the \citet{SH0ES22} statistical uncertainties.
Since that publication, however, the most recent equivalent analysis measured $H_0=73.17\pm0.86$ \kmsMpc{} by improving the first rung of the distance ladder \citep{Breuval2024}. 
Using a reanalyzed SN sample of this more recent distance ladder would also allow us to decrease our systematic uncertainty.

\subsection{Constraints on \texorpdfstring{$H_0$}{H0} and \texorpdfstring{$q_0$}{q0}}
For the SNe Ia, we also perform fits with $q_0$ as a free parameter. For $H_0$, we set a permissive uniform prior of [50.00, 90.00] \kmsMpc\ while for $q_0$ we select a uniform prior of [$-0.58$, $-0.48$]. The $q_0$ prior reflects a generous expansion of the $q_0$ posterior region found using Dark Energy Survey Year 5 (DES Y5) SNe Ia dataset \citep{DES2024}.  Using the SNe Ia calibrators, our resulting $q_0$-free fitted posteriors for the combined TF+FP dataset are shown in Figure \ref{fig:H0q0_corner}. We also show the posterior on $H_0$ and $q_0$ reported by SH0ES \citep{SH0ES22}. From this fit, we find $H_0=73.8_{-1.1} ^{+1.2}$ \kmsMpc\ and $q_0=-0.56_{-0.02} ^{+0.01}$, acknowledging the strong influence on the $q_0$ constraint stemming from the DES Y5 prior. This result is driven by the much larger FP dataset. 
Similar to the findings of \citet{SH0ES22}, the effect on $H_0$ is small; however, the constraint on $q_0$ is not very reliable given the heavily restricted redshift range of the DESI PV sample.

\section{Summary and Conclusion}\label{sec:summary}
The $>100,000$ galaxy distances directly measured by DESI in the first year of observation provide an exciting opportunity to measure precise cosmology in the nearby Universe.
Not only have the distances been used to measure peculiar velocities and therefore the growth rate of structure as part of the DESI PV survey DR1 analysis \citep{DESI_DR1_PV_max_like, DESI_DR1_PV_dens_vel_corr, DESI_DR1_PV_power_spec, DESI_DR1_PV_SN_xcorr}, but the distance and redshift information allow us to measure the recent expansion history.
The statistical power of this sample is already impressive, and set to grow in future data releases.

As the DESI PV tracers offer only relative distance measures, we combine the PV sample with independent, external calibrators in order to convert to absolute distances.
We combine the TF and FP samples to find a global zero-point primarily through the use of Pantheon+ SNe Ia calibrated by the SH0ES three-rung distance ladder.
With absolute distances measured by the calibrated FP and TF relations, along with the redshifts measured by DESI, we constrain the present-day expansion rate of the Universe.
This results in measurements of $H_0$ highly consistent with the Pantheon+/SH0ES measurement, but with greater statistical precision due to the sheer number of galaxies in the DESI DR1 PV sample.
We also examine several other calibrators: masers, SBF and single distances to the Coma cluster measured via SBF and SNe Ia. 
Since the SBF and SNe Ia we use here are both calibrated by Cepheids, we combined them for greater statistical power.

This enabled a factor of ${\sim}6$ improvement to the statistical-only uncertainty and similarly a factor of ${\sim}5$ in total uncertainty.
The improvement is mainly due to the sample size, but also due to the methodological improvements of this analysis.
The most notable of these is the switch from a single calibrator to many, which was enabled in part by the use of a group catalog.
Our method for defining the group catalog was to bring together four different catalogs (2dFGRS, 6dFGS, 2MRS and SDSS) created and analyzed in \citet{Lim2017}, in the absence of a DESI BGS-only group catalog.
In the future, a DESI-only group catalog would have much higher magnitude completeness than any other group catalog available at the time of writing. 
One of the largest systematics of this analysis relates to the group catalog, which requires a correction to account for the difference between grouped and non-grouped galaxies.
However, one caveat is that the source of the mechanism requiring correction has not been confirmed, but it is consistent with the hypothesis of a relatively incomplete group catalog.

When we tested the effect of choices of analysis methods on the recovered constraints on the Hubble parameter, we found that our measurement is robust to the choices made.
The largest change was related to allowing the very low-$z$ (biased) region to be used for calibration, followed by group-related systematics.
Not applying the group correction is a large effect, although it is not unexpected, as removing the correction is tantamount to zero-pointing to a biased mean.
However, even the largest systematics only shift the recovered $H_0$ by at most ${\sim}1\sigma$.

After examining the different calibrators that overlap with DR1, we conclude that supernovae currently remain the best way to calibrate the PV sample.
This is mostly due to the much greater redshift and sky overlap, which will only increase in the future.
This approach, however, still inherits the uncertainty from all previous distance ladder rungs.
There is not yet a detailed study of the TF and FP systematics that we can propagate through our analysis as, \textit{e.g.}, a full covariance matrix. 
However, we showed in our brief study in Sec.~\ref{sec:cosmology} that the DESI PV specific systematics appear small compared to the distance ladder systematics.
If we can control these sources of systematics and increase the overlap with primary calibrators, the DESI PV measurements will be a viable alternative to the traditional SN distance ladder.
We already saw evidence that SBF measurements by themselves could approach the statistical precision of SN calibration.
Currently, most SBF measurements are too local to use in DR1, but DR2 may see a greater overlap or include a very local correction to measured TF and FP distances to remove any potential bias at low-$z$.

Our best constraint on the Hubble parameter comes from SNe Ia alone, giving $H_0=\HSN\pm\HstatSN\;\mathrm{(stat.)}\pm\HtotsystSN\;(\text{syst.})=\HSN\pm\HtotSN$ \kmsMpc{}. 
Our tightest constraint comes from combining SNe and SBF, but the total uncertainty on that measurement is a slight underestimate without a full treatment of SBF covariance. 
Currently, the main limitation of our measurements is the limited number of external calibrators, which should improve with future data.
If we can better control systematics and transition to a distance ladder with fewer rungs, we expect to approach or perhaps even reach a local $H_0$ measurement at the sub-percent level.

\section*{Data Availability}

The data used to create the figures will be made available on Zenodo (DOI: \dataset[10.5281/zenodo.17548130]{https://doi.org/10.5281/zenodo.17548130}) upon acceptance of the papers in this series. Other data products of the DESI DR1 PV Survey will also be made available at \url{https://data.desi.lbl.gov/doc/releases/dr1/}.

\begin{acknowledgements}
This material is based upon work supported by the U.S. Department of Energy (DOE), Office of Science, Office of High-Energy Physics, under Contract No. DE–AC02–05CH11231, and by the National Energy Research Scientific Computing Center, a DOE Office of Science User Facility under the same contract. Additional support for DESI was provided by the U.S. National Science Foundation (NSF), Division of Astronomical Sciences under Contract No. AST-0950945 to the NSF’s National Optical-Infrared Astronomy Research Laboratory; the Science and Technology Facilities Council of the United Kingdom; the Gordon and Betty Moore Foundation; the Heising-Simons Foundation; the French Alternative Energies and Atomic Energy Commission (CEA); the National Council of Humanities, Science and Technology of Mexico (CONAHCYT); the Ministry of Science, Innovation and Universities of Spain (MICIU/AEI/10.13039/501100011033), and by the DESI Member Institutions: \url{https://www.desi.lbl.gov/collaborating-institutions}. Any opinions, findings, and conclusions or recommendations expressed in this material are those of the author(s) and do not necessarily reflect the views of the U. S. National Science Foundation, the U. S. Department of Energy, or any of the listed funding agencies.

The authors are honoured to be permitted to conduct scientific research on I'oligam Du'ag (Kitt Peak), a mountain with particular significance to the Tohono O’odham Nation.
\end{acknowledgements}

\bibliography{references,DESI_supporting_papers}{}
\bibliographystyle{aasjournalv7}

\end{document}